\documentclass[prd,aps,10pt,twocolumn,nofootinbib,showpacs]{revtex4-2}

\usepackage{amssymb,amsmath}
\usepackage{xcolor}
\usepackage[utf8]{inputenc}

\usepackage[export]{adjustbox}

\usepackage{float}
\usepackage{enumitem}
\usepackage{empheq}
\usepackage{graphicx}
\usepackage{framed}
\usepackage[makeroom]{cancel}

\usepackage[makeroom]{cancel}
\usepackage[caption=false]{subfig}
\usepackage[colorlinks=true,
            citecolor=green,
            linkcolor=red,
            filecolor=cyan,
            urlcolor=magenta,
            backref=false]{hyperref}

\newcommand{\dd}{\mbox{d}}

\begin{document}

\title{Microlensing of non-singular black holes at finite size: a ray tracing approach}

\author{Jens Boos}
\email{jens.boos@kit.edu}
\affiliation{Institute for Theoretical Physics, Karlsruhe Institute of Technology, D-76128 Karlsruhe, Germany}

\author{Hao Hu}
\email{hao.hu@student.kit.edu}
\affiliation{Institute for Theoretical Physics, Karlsruhe Institute of Technology, D-76128 Karlsruhe, Germany}

\date{December 18, 2025}

\begin{abstract}

We study the gravitational microlensing of various static and spherically symmetric non-singular black holes (and horizonless, non-singular compact objects of similar size). For pointlike sources we extend the parametrized post-Newtonian lensing framework to fourth order, whereas for extended sources we develop a ray tracing approach via a simple radiative transfer model. Modelling non-relativistic proper motion of the lens in front of a background star we record the apparent brightness as a function of time, resulting in a photometric lightcurve. Taking the star radius to smaller values, our numerical results approach the theoretical predictions for point-like sources. Compared to the Schwarzschild metric in an otherwise unmodified lensing geometry, we find that non-singular black hole models (and their horizonless, non-singular counterparts) both at finite size and for point lenses tend to feature larger magnifications in microlensing lightcurves.

\end{abstract}

\maketitle

\section{Introduction}

The bending of light has played a pivotal role in establishing general relativity as a unified theory of gravity and relativity \cite{Will:2014kxa}. Direct observation of supermassive black holes and their accretion disk has given first images of black holes \cite{EventHorizonTelescope:2019dse}. In the context of time domain astronomy, bending of light has been utilized to identify the passage of otherwise invisible, rogue compact astrophysical objects via so-called microlensing events \cite{Paczynski:1985jf} (for a recent review, see Ref.~\cite{Abrams:2023vzw} and reference therein).

On the theoretical side there has been much activity in the study of compact astrophysical objects, with an increasing focus on black holes. Importantly, one of the key problems of black holes as described by general relativity---namely, that their interior contains spacetime singularities---has not been resolved. Conversely, a large community has begun to develop phenomenologically motivated black holes, called ``regular black holes'' or ``non-singular black holes,'' which are typically accompanied by a length scale parameter $\ell > 0$ that plays the role of a regulator. While some models can be motivated by physical principles or shown to solve exact Lagrangians, these models are typically not viewed as fundamental predictions of an underlying quantum theory of gravity, but rather as toy models that parametrize possible deviations from the black hole metrics of general relativity \cite{Frolov:2016pav,Carballo-Rubio:2025fnc}. Besides improving the mathematical and conceptual consistency of such models, an important goal is to robustly explore the phenomenological implications of such non-singular black hole models and to identify experimental methods suitable to constrain such models.

For a light ray trajectory in a black hole spacetime, the majority of gravitational deflection occurs in the direct vicinity of the gravitational radius \cite{Narayan:1996ba,Wambsganss:1998gg,Falcke:1999pj,Perlick:2004tq,Cunha:2018acu,Gralla:2019xty,Perlick:2021aok}. For this reason, the deflection of light is rather sensitive to horizon-scale modifications and is under active investigation (see, e.g., \cite{Amarilla:2010zq,Atamurotov:2013sca,Wei:2013kza,Tsukamoto:2014tja,Perlick:2015vta,Abdujabbarov:2016hnw,Tsukamoto:2017fxq,Konoplya:2019sns}) via presently available strong-lensing data stemming from the vicinity of supermassive black holes. However, such astrophysically active regions impede the identification of signatures of new physics \cite{Mizuno:2018lxz,Volkel:2020xlc,Vagnozzi:2022moj,Broderick:2023jfl}.

Free floating ``rogue'' astrophysical objects are typically not surrounded by accretion disks, and their velocities, as typical solar-system-scale escape velocities, are non-relativistic, such that their motion can be modeled without the necessity of taking relativistic effects into account. While otherwise practically invisible, such objects can lens the light of a background star in a process called ``microlensing.'' The resulting increase in brightness of the distant, luminous source under the passage of the otherwise invisible, compact object, is then captured in a photometric lightcurve. In that process, the order of magnitude for the angular scale of the lensing process is given by the Einstein angle $\vartheta_\text{E}$ which is assumed small,
\begin{align}
\vartheta_\text{E} = \sqrt{\frac{4GM d_\text{ls}}{c^2\,d_\text{ol}d_\text{os}}} \ll 1 \, ,
\end{align}
and the lensed image cannot be resolved. Here, $M$ is the mass of the lens, $d_\text{ls}$ is the distance from the lens plane to the source plane, $d_\text{ol}$ is the distance from the observer plane to the lens plane, and $d_\text{os}$ is the distance from the observer plane to the source plane. All planes are orthogonal to the observer-star axis (the optical axis in this lensing system). For stellar masses and distances of kiloparsecs, the resulting angle is $\vartheta_\text{E} \in \mathcal{O}(10^{-9})$. Fixing as an optical axis the line of sight between a fixed background star (the source) and the observer, we write the angle subtended by the lens as $\beta \vartheta_\text{E}$ where $\beta$ is a dimensionless angle. Expressed in that variable, the magnification of the star is then given, to leading order, by
\begin{align}
\mu(\beta) \approx \frac{2+\beta^2}{\beta\sqrt{4+\beta^2}} \, .
\end{align}
The proper motion of the lens induces a time dependence in $\beta$ and hence the above magnification becomes a function of time. Typical distance scales, lens masses, optical depths, and proper motion of lenses vary from scenario to scenario---historically, microlensing has been proposed as an avenue to search for black holes or similarly dark and compact objects as potential dark matter candidates, but it has also been successful in locating exoplanets, which is is mirrored in the multitude of past, present, and planned microlensing surveys such as OGLE, MOA, KMTNet, Kepler, and Roman (focused on the galactic bulge, looking for stellar and planetary microlensing) as well as MACHO, EROS, SuperMACHO, and Subaru HSC (focusing on the galactic halo, the Magellanic clouds, or Andromeda, looking for dark matter candidates).

Because microlensing signatures stem from isolated compact gravitational objects, photometric lightcurves are a comparatively straightforward avenue to model deviations from general relativity. A general perturbative formalism to describe lensing in static, spherically symmetric spacetimes has been developed by Keeton and Petters \cite{Keeton:2005jd}, and while strong lensing and black hole shadows have received considerable attention in the context of non-singular black hole models \cite{Amarilla:2010zq,Atamurotov:2013sca,Wei:2013kza,Tsukamoto:2014tja,Perlick:2015vta,Abdujabbarov:2016hnw,Tsukamoto:2017fxq,Konoplya:2019sns}, to the best of our knowledge microlensing and photometric lightcurves of non-singular compact objects have not yet been described in detail (see, however, Refs.~\cite{Gao:2022gbn,Liu:2022lfb,Gao:2022cds,Gao:2023sla,Verma:2023hes,Gao:2023ipv,Li:2025mqx}). In this paper, we will follow a dual approach:
\begin{itemize}
\item We first extend the parametrized post-Newtonian (PPN) framework by Keeton and Petters from the third to the fourth order, such that it captures the first PPN deviation of the Hayward metric from the Schwarzschild metric. This is sufficient to describe the microlensing phenomenology for pointlike sources.
\item In a second step, we will implement a numerical ray tracing code for static, spherically symmetric metrics supplemented by a simple radiative transfer model. By integrating the brightness recorded by the detector screen, we can numerically derive the expected lightcurve in a wide range of scenarios for extended sources.
\end{itemize}
This approach guarantees that the correct method can be utilized depending on the size of the source, and we will moreover verify that the two methods approach each other in a reasonable manner.

This paper is organized as follows. In Sec.~\ref{sec:bh-models} we will briefly discuss general features of non-singular black hole models, and then describe the models considered in this paper in more detail. In Sec.~\ref{sec:gravitational-lensing} we derive the weak gravitational lensing angle and then extend the PPN formulation to fourth order, arriving at an analytical perturbative formula for the microlensing lightcurve applicable to pointlike sources. We then describe a numerical ray tracer in Sec.~\ref{sec:ray-tracing}, and in Sec.~\ref{sec:strong} we test this ray tracer via a simple accretion disk model around a non-singular black hole in the strong-lensing regime. We turn towards numerical microlensing in Sec.~\ref{sec:microlensing} by modeling the motion of a rogue black hole that intersects the line of sight between the observer and a distant but extended background star light source. The ray traced intensities for each pixel are then integrated to yield the lightcurve for each scenario, allowing us to quantify the deviations between the Schwarzschild lightcurve prediction and various non-singular black hole models. We also extract the lightcurves of compact horizonless objects, which are obtained from increasing the regularization parameter $\ell$ beyond a critical value. By reducing the size of the light source, we show that the numerical results converge to the PPN expressions valid only for pointlike sources. We conclude our findings in Sec.~\ref{sec:conclusion} and in appendix \ref{app:ppn} we list complete PPN formulas to fourth order, and the numerical information for the lightcurve simulations as well as additional ray traced images are summarized in appendix \ref{app:images}.

\section{Non-singular black hole models}
\label{sec:bh-models}

In the context of this work, we focus our attention to the case of static, spherically symmetric metrics of the form
\begin{align}
\label{eq:metric}
\dd s^2 = - A(r) \dd t^2 + \frac{\dd r^2}{B(r)} + C(r)^2 \dd\Omega^2 \, .
\end{align}
Here, $\dd\Omega^2 = \dd\theta^2 + \sin^2\theta\,\dd\varphi^2$ is the usual angular surface element, and $A(r)$, $B(r)$, and $C(r)$ are general functions subject only to the asymptotic flatness contraints
\begin{align}
\label{eq:asymptotically-flat}
A(r \rightarrow \infty) = B(r \rightarrow \infty) = 1 \, , \quad
C(r \rightarrow \infty) = r \, .
\end{align}
The metric \eqref{eq:metric} can be further simplified assuming that $C(r)$ is a monotonous function. In that case, one can define $C(r)$ as a new radial variable, which then changes the $g_{rr}$ component in the new metric, resulting in simply setting $C=1$ in the above metric. However, we opted to keep the slightly more general form for ease of application: namely, while for all of our examples we have $A(r) = B(r)$, we will also consider a non-singular black hole model containing wormhole-like features mediated via a non-standard parametrization with $C(r) \not= r$. By keeping the metric this flexible, we also hope that this work will be more useful to other researchers since initial transformations of radius variables are not necessary.

Non-singular black hole models are phenomenologically motivated, parametrized metrics that reduce to exact black hole solutions of general relativity in the limiting procedure of vanishing regulator parameter \cite{Frolov:2016pav,Carballo-Rubio:2025fnc}. In the present context, we shall only utilize metrics with a single regulator $\ell > 0$ with the physical dimension of length. Focussing now on spherically symmetric models, let us address a few common features.

\begin{itemize}[leftmargin=12pt]
\item Spacetime curvature scalars, which are singular at $r=0$ for the Schwarzschild black hole, are finite everywhere. Their maximum values schematically take the following form:
\begin{align}
\text{curvature} \sim \frac{1}{\ell^2} ~ \text{or} ~ \frac{GM}{\ell^3} \, .
\end{align}
The first form is independent of the black hole mass, and is encountered in models implementing the so-called ``limiting curvature condition'' that demands not only finite curvature everywhere in space, but also finite curvature as the mass $M$ increases to infinity, thereby implementing a universal bound on spacetime curvature. The second form is more agnostic and can be deduced from dimensional analysis. We note in passing that the location of maximum curvature is not always $r=0$ but may be instead located at a small to intermediate-value radius.
\item The geometry at the coordinate origin $r=0$ is smooth, which implies
\begin{align}
\begin{split}
\label{eq:core}
A(r \rightarrow 0) \sim 1 - a_{0,2} r^2 + \mathcal{O}(r^3) \, , \\
B(r \rightarrow 0) \sim 1 - b_{0,2} r^2 + \mathcal{O}(r^3) \, ,
\end{split}
\end{align}
for some constants $a_{0,2}$ and $b_{0,2}$. Note the absence of a linear term in the near-origin expansion, since that would induce a conical singularity. The Ricci scalar for such a metric, assuming that $C(r) = r + c_0$, to leading order is $R = 2a_{0,2} + \mathcal{O}(r)$, implying that the sign of $a_{0,2}$ sets the sign of local curvature at the origin. This implies that $a_{0,2} > 0$ features positive curvature, which is also referred to as ``de\,Sitter like,'' and $a_{0,2} = 0$ instead leads to vanishing scalar curvature, which is called ``Minkowski like.''
\item The black hole horizon for the metric \eqref{eq:metric} is given by $B(r) = 0$. However, depending on the size of the regulator $\ell > 0$, this equation has different characteristics. For vanishing $\ell$ one recovers the horizon location of the Schwarzschild metric, but for larger $\ell > 0$ one instead finds both an outer horizon and an inner horizon. Often, beyond some critical value $\ell \geq \ell_\star$ no horizon exists. Since the only other scale in the system is the black hole mass $M$, the condition for the existence of a black hole horizon can then be recast into the form
\begin{align}
M > M_\star (\ell_\star) \, ,
\end{align}
which is also referred to as a ``mass gap'' \cite{Frolov:2015bta}. Recently this notion was extended to an entire ``band structure'' for mass-dependent regulators \cite{Boos:2023icv,Asmanoglu:2025agc}.
\item The lightlike effective potential for the metric \eqref{eq:metric} is
\begin{align}
V_\text{eff} = \frac{B}{A} - \frac{b^2 B}{C^2} \, ,
\end{align}
where $b$ is the impact parameter of the null geodesic. The term $B/A$ is usually absent from an effective null potential since $A=B$ in the case of the Schwarzschild metric, and, in fact, the metrics we discuss in this paper also satisfy $A=B$. More interesting is the second term which is proportional to B. In the singular Schwarzschild case we see that for small distances the angular-momentum repulsion dominates this potential. However, in non-singular geometries the function $B$ is finite at small radii, which means that the potential will exhibit an additional inflection point, called an inner light ring \cite{Cunha:2017qtt,Eichhorn:2022oma,Cunha:2022gde,DiFilippo:2024ddg}.

\item Since the surface $r=0$ is regular, the radial variable $r$ may be analytically continued beyond that surface. However, not all non-singular black hole models allow such transformations, leading to significant distinctions in terms of geodesic completeness \cite{Bambi:2016wdn,Carballo-Rubio:2019fnb,Zhou:2022yio}.

\end{itemize}
The last three properties have important phenomenological consequences for the propagation of light rays, since they directly affect the existence of horizons, and---in the absence thereof---the possibility of light rays ``vanishing'' into the space beyond $r=0$. However, in the context of microlensing the latter light rays form an infinitesimally small subset in the class of light rays contributing to observed brightness levels of luminous background objects, which is why we will not consider this issue further in the present work. We will, however, carefully distinguish geometries with and without black hole horizons, as they lead to drastically different phenomenologies due to the existence of additional light rings.

\subsection{Benchmark scenarios}

Our numerical study is designed to determine qualitative differences of static spherically symmetric non-singular black hole geometries as compared to the Schwarzschild metric of general relativity, working at a fixed reference mass. To focus the discussion somewhat, we will limit our considerations to three cases of non-singular black holes:
\begin{itemize}[leftmargin=12pt]
\item The Hayward metric, an example of a de\,Sitter core geometry with positive scalar curvature \cite{Hayward:2005gi},
\item the Minkowski core metric (with vanishing scalar curvature at its core) \cite{Simpson:2019mud},
\item and the Simpson--Visser metric (with a non-standard wormhole-type core) \cite{Simpson:2018tsi}.
\end{itemize}
We will refer to these three models as H, M, and W, respectively; their metrics are listed in Table \ref{table:metrics}. For each metric we will consider two scenarios:
\begin{itemize}[leftmargin=12pt]
\item The subcritical case, where the regulator scale is below the critical value, $\ell < \ell_\star$, such that a black hole horizon exists. We choose the value of the regulator close to $\ell_\star$ to simplify the numerical capture of the observational signatures of such a regulator. This is theoretically motivated by the recently discussed mass-dependent regulators, where one of us argued that mass-dependent regulators---as expected from limiting curvature considerations---can lead to percent-level deviations of the metric at the horizon scale for astrophysically relevant stellar-mass black holes \cite{Boos:2023icv}.
\iffalse \item The ``bumpy'' case, where we add to the subcritical metric a parametrized bump outside the horizon at $r=r_\text{h}$. We implement this by adding to the metric functions $A$ and $B$ the function
\begin{align}
\frac{c_1}{\left[1 + c_2 (r-c_3 r_\text{h})^2\right]^{c_4}} \, .
\end{align}
We ensure that this bump is not large enough to induce new horizons; instead, it will serve as a stand-in to probe the sensitivity of our numerical studies to localized variations in the gravitational field. \fi
\item The supercritical case, where the regulator scale exceeds the critical value, $\ell > \ell_\star$, resulting in the geometry of a compact horizonless object.
\end{itemize}
We denote the subcritical case with ``-'' and the supercritical case with ``+,'' resulting in the seven scenarios $\{S, H_-, H_+, M_-, M_+, W_-, W_+\}$, where $M_+$, for example, stands for the supercritical Minkowski core scenario, and so on. In principle these studies can be extended to an arbitrary number of non-singular models, but the above set corresponds to a reasonable compromise.

\begin{table*}[!htb]
\begin{tabular}{lclclclclclcr} \hline \hline
Abbreviation &~& Name &~& $A(r) = B(r)$ &~& $C(r)$ &~& Horizon condition &~& Ref. \\[3pt] \hline
&&&&&& \\[-8pt]
S && Schwarzschild && $\displaystyle 1 - \frac{2GM}{r}$ && $\displaystyle r$ && $\displaystyle GM > 0$ \\[12pt]
H && Hayward && $\displaystyle 1 - \frac{2GM}{r} \frac{r^3}{r^3 + 2GM \ell^2} $ && $\displaystyle r$  && $\displaystyle 0 \leq \frac{\ell}{2GM} \leq \frac{2}{3\sqrt{3}} \approx 0.3849$ && \cite{Hayward:2005gi} \\[12pt]
M && Minkowski core && $\displaystyle 1 - \frac{2GM}{r} e^{-\ell/r}$ && $\displaystyle r$ && $\displaystyle 0 \leq \frac{\ell}{2GM} \leq \frac{1}{e} \approx 0.3679$ && \cite{Simpson:2019mud} \\[12pt]
W && Simpson--Visser && $\displaystyle 1 - \frac{2GM}{\sqrt{r^2+\ell^2}} $ && $\displaystyle \sqrt{r^2+\ell^2}$ && $\displaystyle 0 \leq \frac{\ell}{2GM} \leq 1$ && \cite{Simpson:2018tsi} \\[12pt]
\hline \hline
\end{tabular}
\caption{We list the three non-singular black hole metric functions under consideration in this paper, where the metric is given by $\dd s^2 = -A(r)\dd t^2 + \dd r^2/B(r) + C(r)^2\dd\Omega^2$. For each metric, we list the conditions on the regulator for the existence of a black hole horizon.}
\label{table:metrics}
\end{table*}

\section{Weak gravitational lensing}
\label{sec:gravitational-lensing}

\begin{figure}[!htb]
\centering
\includegraphics[width=0.42\textwidth]{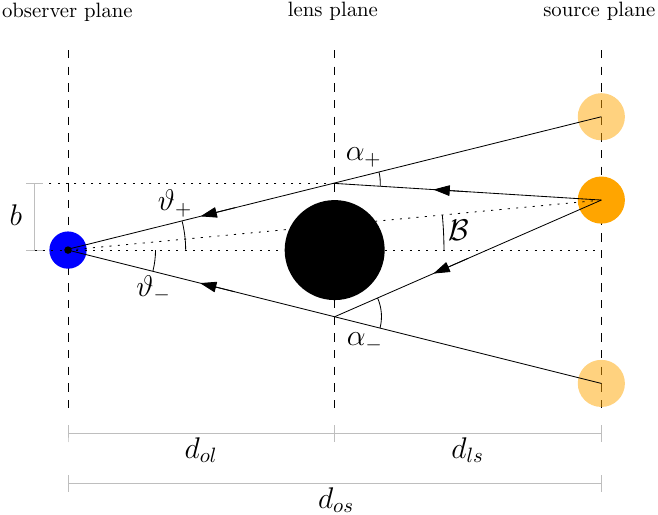}
\caption{Visualization of the lensing setup utilized in this paper. For $\mathcal{B} \not=0$ the upper (lower) image is denoted by a plus (minus). Generally, we utilize $\vartheta$ for the image positions, $\mathcal{B}$ for the source positions, and $\alpha$ for the total deflection angle.}
\label{fig:lensing}
\end{figure}

In this section, we briefly discuss notions encountered in weak gravitational lensing. In order to compare our numerical results with the theoretical expectation, we will first focus on the leading-order expressions that can be derived from first principles, before, in a second part, extending known results from the parametrized post-Newtonian (PPN) framework from third to fourth order. In all of this, we will limit our considerations to the weak deflection angle $\alpha$, the Einstein angle $\vartheta_\text{E}$, the lensed image positions $\vartheta_\pm$, as well as the total magnification $\mu_\text{tot}$ and the resulting light curve in presence of a moving lens. For our convention of the lensing geometry refer to Fig.~\ref{fig:lensing}.

\subsection{Weak lensing angle}
Defining the impact parameter for a null geodesic as $b \equiv L/E$ with $E = A(r)\dot{t}$ and $L = C(r)^2\dot{\varphi}$, without loss of generality we set $E=1$ and find for the exact deflection angle
\begin{align}
\label{eq:deflection}
\alpha = 2b \!\! \int\limits_{r_0(b)}^\infty \frac{\dd r}{C^2} \left[ \left( \frac{1}{A} - \frac{b^2}{C^2} \right) B \right]^{-1/2} - \pi \, .
\end{align}
While we will utilize this expression to derive the weak deflection angle perturbatively in the following, at this stage it is computationally convenient (and conceptually interesting) to utilize topological methods in connection with the Gauss--Bonnet theorem \cite{Gibbons:2008rj} applied to the optical metric in the equatorial plane (obtained from $\dd s^2 = 0$),
\begin{align}
\dd t^2 = \frac{1}{A(r)} \left[ \frac{\dd r^2}{B(r)} + C(r)^2\dd\varphi^2 \right] \, .
\end{align}
The bending angle can then be determined as  \cite{Gibbons:2008rj}
\begin{align}
\alpha = (-1) \int\!\!\int \dd S \, K \, , \quad \dd S = \frac{C}{\sqrt{A^2 B}} \, \dd r \, \dd\varphi \, ,
\end{align}
where for large distances we can parametrize the light ray as $r = b/\sin\varphi$ such that
\begin{align} 
\label{eq:approx-gb}
\alpha \approx \frac12 \, \int\limits_0^\pi \dd\varphi \int\limits_{b/\sin\varphi}^\infty \frac{ R \, C \, \dd r}{\sqrt{A^2 B}} \, ,
\end{align}
where we have substituted the two-dimensional identity $K = (-1/2)R$ between the Gauss and Ricci curvature, and we calculate
\begin{align}
\begin{split}
R &= -\frac{B A'^2}{A} + A' \left( \frac{B'}{2} + \frac{B C'}{C} \right) + B A'' \\
&\hspace{11pt} - \frac{A( B'C' + 2 B C'' )}{C} \, .
\end{split}
\end{align}
Substituting $A=B=1-2GM/r$ and $C=r$ one quickly recovers the well-known general relativistic weak deflection angle,
\begin{align}
\alpha = \frac{4GM}{b} + \mathcal{O} \left( \frac{GM}{b} \right)^3 \, ,
\end{align}
which we shall denote as $\alpha_\text{GR} \equiv 4GM/b$ for convenience. We note that application of this method to higher orders in $GM/b$ will yield results incompatible with a consistent expansion scheme, since the approximate relation $r = b/\sin\varphi$ utilized in Eq.~\eqref{eq:approx-gb} is only true to leading order.

\iffalse \begin{align}
\begin{split}
\alpha_\text{H} &= \frac{4GM}{b} + \frac{3 (GM)^2\pi}{4b^2} + \frac{8(GM)^3}{b^3} \\
&\hspace{11pt}+ \frac{75(GM)^4\pi}{64b^4} - \frac{15 (GM)^2\pi\ell^2}{b^4}
\end{split}
\end{align} \fi

\subsection{Fourth-order PPN expressions}

While the deflection angle is solely a function of the impact parameter $b$, in realistic scenarios the properties of a lensing system depend on various distance scales that we visualize in Fig.~\ref{fig:lensing}. As is customary in the case of asymptotically flat spacetimes, we visualize the system as embedded in three-dimensional Euclidean space. We denote the impact parameter of a lensed light ray as $b$, and we denote the distance from the observer to the lens as $d_\text{ol}$, the distance from lens to the star as $d_\text{ls}$, and the total distance between the observer and the star as $d_\text{os}$. Then, utilizing the notion of an angular-diameter distance, we can read off the lens equation \cite{Narayan:1996ba,Wambsganss:1998gg}
\begin{align}
\label{eq:lensing}
d_\text{os} \tan\mathcal{B} = d_\text{os} \tan\vartheta - d_\text{ls} \, \left[ \tan\vartheta + \tan \left( \alpha - \vartheta \right) \right] \, ,
\end{align}
which, upon linearization, becomes the familiar
\begin{align}
d_\text{os} \mathcal{B} \approx d_\text{os} \vartheta - d_\text{ls} \alpha \, .
\end{align}
The above relation holds for both image positions $\vartheta = \vartheta_\pm$, where we then also set $\alpha = \alpha_\pm$ to account for the different bending of light in the presence of $\mathcal{B} \not=0$. In the special case of rotation symmetry around the optical axis between observer, lens, and star (that is, when $\mathcal{B}=0$), we have $\vartheta_+ = \vartheta_-$. In this case the source is mapped to a circle of characteristic angle $\vartheta_E$ and radius $r_E$, called Einstein angle and Einstein radius, respectively:
\begin{align}
\vartheta_\text{E} &= \sqrt{b\,\alpha_\text{GR} \, \frac{d_\text{ls}}{d_\text{ol}d_\text{os}}} = \sqrt{\frac{4GM\,d_\text{ls}}{d_\text{ol}d_\text{os}}} \, , \\
r_\text{E} &= d_\text{ol} \tan\vartheta_E = d_\text{os} \tan \sqrt{\frac{4GM\,d_\text{ls}}{d_\text{ol}d_\text{os}}} \, .
\end{align}
It is now convenient to normalize all angles in terms of the Einstein angle $\vartheta_\text{E}$, and we define the dimensionless
\begin{align}
\beta \equiv \frac{\mathcal{B}}{\vartheta_\text{E}} \, , \quad
\theta \equiv \frac{\vartheta}{\vartheta_\text{E}} \, .
\end{align}
We are now ready to derive a parametrized-post Newtonian (PPN) expression for the bending angle, extending the results of Ref.~\cite{Keeton:2005jd} from third to fourth order. This is necessary since the Hayward metric has radial subleading contributions at fourth order,\footnote{The appearance of mass-dependent regulators, such as the $GM\ell^2$ in the denominator of the Hayward metric, does not spoil the PPN expansion scheme, as long as $\ell/(GM)$ is at most around $\mathcal{O}(1)$. In the present context, this will always be the case.}
\begin{align}
\begin{split}
f_\text{H} &= 1 - \frac{2GM}{r} \frac{r^3}{r^3 + 2GM\ell^2} \\
&\approx 1 - \frac{2GM}{r} + \frac{4(GM)^2\ell^2}{r^4} + \mathcal{O}\left( \frac{(GM)^3 \ell^4}{r^7} \right) \, .
\end{split}
\end{align}
In order to allow for a straightforward comparison of our results to that found in the literature, we will deviate in this subsection from Eq.~\eqref{eq:metric} and instead use
\begin{align}
\dd s^2 = -A(\tilde{r}) \dd t^2 + \tilde{B}(\tilde{r}) \dd \tilde{r}^2 + \tilde{r}^2 \dd\Omega^2 \, ,
\end{align}
where we read off
\begin{align}
\label{eq:c-trafo}
C(r) = \tilde{r}(r) \, , \quad
\frac{\dd r^2}{B(r)} = \tilde{B}(\tilde{r}) \dd \tilde{r}^2 \, ,
\end{align}
which can be used to convert between the two metrics. Then, utilizing the expression (note the minus sign)
\begin{align}
\phi = -\frac{GM}{\tilde{r}} \, ,
\end{align}
we define the PPN expansion coefficients $a_i$ and $b_i$ to fourth order via
\begin{align}
\begin{split}
\label{eq:ppn-def}
A &= 1 + 2a_1 \phi + 2 a_2 \phi^2 + 2 a_3 \phi^3 + 2 a_4 \phi^4 + \mathcal{O}(\phi^5) \, , \\
\tilde{B} &= 1 - 2b_1 \phi + 4b_2 \phi^2 - 8 b_3 \phi^3 + 16 b_4 \phi^4 + \mathcal{O}(\phi^5) \, ,
\end{split}
\end{align}
and the values for the metrics under consideration in this paper can be found in Table~\ref{table:ppn-coefficients} in the appendix. It is helpful at this point to express $r_0$ (the point of closest approach of a null geodesic to the black hole) in terms of the impact parameter $b$ using the above expansion. To do so, we first expand
\begin{align}
b = \frac{r_0}{\sqrt{A(r_0)}}
\end{align}
as a power series in $GM/r_0$ and then define
\begin{align}
r_0 = b \left[ 1 + \sum\limits_{n=1}^4 c_n \left(\frac{GM}{b}\right)^n + \mathcal{O}\left(\frac{GM}{b}\right)^5 \right] \, .
\end{align}
Substituting the latter into the power series for $b$ we demand consistency and thereby read off the coefficients
\begin{align}
\label{eq:r0-via-b}
c_1 &= -a_1 \, , \\
c_2 &= \frac{2a_2 - 3 a_1^2}{2} \, , \nonumber \\
c_3 &= - 4 a_1^3 + 4 a_1 a_2 - a_3 \, , \nonumber \\
c_4 &= \frac18 \left( -105 a_1^4 + 140 a_1^2 a_2 - 20 a_2^2 - 40 a_1 a_3 + 8 a_4 \right) \, . \nonumber
\end{align}
This expansion is very convenient since it expresses the point of closest approach to the impact parameter $b$ and the ADM mass $M$ (and, possibly, the regulator $\ell$).

We can now turn to the deflection of light \eqref{eq:deflection}, which, expressed in terms of the variable $x = r_0/r$, can be written concisely as
\begin{align}
\alpha = 2 \int\limits_0^1 \dd x \sqrt{\frac{A\tilde{B}}{\left(\frac{r_0}{b}\right)^2 - A x^2}} - \pi \, .
\end{align}
Defining the perturbative quantity $h = GM/r_0$ the integrand can be expanded and then integrated order by order. The result will depend on $r_0$, which is a coordinate-dependent quantity, which is why it is more convenient to express $r_0$ in terms of $b$ via \eqref{eq:r0-via-b}. We define
\begin{align}
\label{eq:ppn-alpha}
\alpha = \sum \limits_{n=1}^4 A_n \left( \frac{GM}{b} \right)^n + \mathcal{O}\left( \frac{GM}{b} \right)^5 \, ,
\end{align}
and find
\begin{align}
A_1 &= 2(a_1+b_1) \, , \\
A_2 &= \left( 2a_1^2 - a_2 + a_1 b_1 - \frac14 b_1^2 + b_2 \right) \, \pi \, , \\
A_3 & = \frac{70}{3} a_1^3 - 20 a_1 a_2 + 4 a_3 + 10 a_1^2 b_1 - 4 a_2 b_1 \\
&\hspace{11pt} - 2 a_1 b_1^2 + \frac23 b_1^3 + 8 a_1 b_2 - \frac83 b_1 b_2 + \frac{16}{3}b_3 \, , \nonumber \\
A_4 &= \bigg( 30 a_1^4 - 36 a_1^2a_2 + \frac92 a_2^2 + 9 a_1 a_3 - \frac32 a_4 \\
&\hspace{11pt} + 12 a_1^3 b_1 - 9 a_1 a_2 b_1 + \frac32 a_3 b_1 - \frac94 a_1^2 b_1^2 + \frac34 a_2 b_1^2 \nonumber \\
&\hspace{11pt} + \frac34 a_1 b_1^3 - \frac{15}{64} b_1^4 + 9 a_1^2 b_2 - 3 a_2 b_2 - 3 a_1 b_1 b_2 \nonumber \\
&\hspace{11pt} + \frac98 b_1^2 b_2 - \frac34 b_2^2 - \frac32 b_1 b_3 + 6 a_1 b_3 + 3 b_4 \bigg) \pi \, . \nonumber 
\end{align}
The expressions $A_i$ will form the basis for many comparisons of our numerical ray traced data to theoretical perturbative expressions.

In order to apply this formalism to the lensing equation \eqref{eq:lensing}, we assume that the angular size of the lens (that is, the black hole) is much smaller than the Einstein angle, arriving at a small expansion parameter
\begin{align}
\label{eq:epsilon}
\epsilon = \frac{\vartheta_\bullet}{\vartheta_\text{E}} = \sqrt{ \frac{d_\text{ol} d_\text{os}}{4GM\,d_\text{ls}} } \arctan\left( \frac{GM}{d_\text{ol}} \right) ~ \ll 1 \, .
\end{align}
We can then postulate the following form of the image position,
\begin{align}
\label{eq:theta-expansion}
\theta = \theta_0 + \theta_1 \epsilon + \theta_2 \epsilon^2 + \theta_3  \epsilon^3  + \mathcal{O}\left(\epsilon^4\right) \, .
\end{align}
Noting that $b = d_\text{ol}\sin\vartheta$, we can insert the above relations into the lensing equation \eqref{eq:lensing} for the general form of the deflection angle \eqref{eq:ppn-alpha} to eliminate the expansion factors $GM/b$, arriving at an expansion in $\epsilon/\theta_0$ instead:
\begin{align}
\alpha &= A_1 \left( \frac{\epsilon}{\theta_0} \right) + (A_2-A_1\theta_1) \, \left(\frac{\epsilon}{\theta_0}\right)^2 \\
&\hspace{11pt}+ \left[ \frac83 A_1 d^2\theta_0^4 A_3 -2 A_2 \theta_1 + A_1 \theta_1^2 - A_1 \theta_0 \theta_2 \right] \left(\frac{\epsilon}{\theta_0}\right)^3 \nonumber \\
&\hspace{11pt}+ \mathcal{O} \left(\frac{\epsilon}{\theta_0}\right)^4 \, , \nonumber
\end{align}
with $d \equiv d_\text{ls}/d_\text{os}$. Inserting the above results back into the lensing equation \eqref{eq:lensing}, we need to recover a null result for consistency, which we can then use to express the image positions $\theta_i$ in terms of the constants $A_i$. We find
\begin{align}
\beta &= \theta_0 - \frac{A_1}{4\theta_0} \, , \label{eq:beta} \\
\theta_1 &= \frac{A_2}{A_1 + 4\theta_0^2} \, , \nonumber \\
\theta_2 &= \frac{1}{3\theta_0(A_1 + 4\theta_0^2)} \bigg( A_1^3 + 3 A_3 - 12 A_1^2 d \theta_0^2 + 64 d^2 \beta^3 \theta_0^3 \nonumber \\
&\hspace{11pt}  + 56 A_1 d^2 \theta_0^4 - 64 d^2 \theta_0^6 - 6 A_2 \theta_1 + 3 A_1 \theta_1^2
\bigg) \, , \nonumber \\
\theta_3 &= \frac{1}{3\theta_0^2(A_1 + 4\theta_0^2)} \bigg( 3 A_1^2 A_2 + 3 A_4 - 24 A_1 A_2 d \theta_0^2 \nonumber \\
&\hspace{11pt} + 64 A_2 d^2 \theta_0^4 - 3 A_1^3 \theta_1 - 9 A_3 \theta_1 + 12 A_1^2 d \theta_0^2 \theta_1 \nonumber \\
&\hspace{11pt} + 56 A_1 d^2 \theta_0^4 \theta_1 - 192 d^2 \theta_0^6 \theta_1 + 9 A_2 \theta_1^2 \nonumber \\
&\hspace{11pt} - 3 A_1 \theta_1^3 - 6 A_2 \theta_0 \theta_2 +  6 A_1 \theta_0 \theta_1 \theta_2 \bigg) \, . \nonumber
\end{align}
In some cases it may be convenient to express $\beta$ in terms of $\theta_0$ and to substitute the explicit results for $\theta_1$, $\theta_2$, and $\theta_3$, yielding explicit expressions for all angles $\theta_i$ that solely depend on $\theta_0$, but for the sake of brevity we will not list them here but instead defer to appendix \ref{app:ppn}. We do wish to emphasize, however, that Eq.~\eqref{eq:beta} relates $\beta$ to two image positions,
\begin{align}
\theta_0^\pm = \frac12 \left( \beta \pm \sqrt{\beta^2 + A_1} \right) \, ,
\end{align}
which then implies
\begin{align}
\theta_0^+ \theta_0^- = -\frac{A_1}{4} \, ,
\end{align}
which is a useful relation when desiring to express one image position $\theta_0^\pm$ via the other $\theta_0^\mp$. In our conventions, cf.~Fig.~\ref{fig:lensing}, the relative sign between the two image positions stems from the identification of the optical axis as the horizontal in the embedding diagram wherein all angles are defined, and it indicates that the two images are on opposite sides of the horizontal axis, as expected.

With the image positions now given as a perturbative solution to the lensing equation, we can define the magnification $\mu$ (for spherically symmetric lenses) as \cite{Narayan:1996ba,Wambsganss:1998gg}
\begin{align}
\label{eq:mu-expansion}
\mu \equiv \mu(\theta) = \left[ \frac{\sin\mathcal{B}}{\sin\theta} \frac{\dd \mathcal{B}}{\dd \theta} \right]^{-1} \, ,
\end{align}
where we will again expand
\begin{align}
\mu = \mu_0 + \mu_1 \epsilon + \mu_2 \epsilon^2 + \mu_3 \epsilon^3 + \mathcal{O}\left( \epsilon^4 \right) \, .
\end{align}
Following the same conceptual steps as above, we express the solution $\theta$ through the expansion \eqref{eq:theta-expansion} and read off the coefficients $\mu_i$, finding
\begin{align}
\mu_0 &= \frac{16\theta_0^2}{A_1^2 - 16\theta_0^2} \, , \\
\mu_1 &= - \, \frac{16 A_2 \theta_0^3}{(A_1 + 4\theta_0^2)^3} \, ,
\end{align}
and the results for $\mu_2$ and $\mu_3$ can be found in appendix \ref{app:ppn}. The total magnification is defined as the sum of the absolute values of the two image magnifications,
\begin{align}
\mu_\text{tot} = |\mu( \theta_0^+ )| + |\mu (\theta_0^- )| \, ,
\end{align}
where we have made use here of the fact that we expressed all coefficients purely in terms of $\theta_0$, and care has to be taken to ensure the positivity of the individual contributions. This can be easily verified, though, by computing the magnifications explicitly for the Schwarzschild geometry. We can then use Eq.~\eqref{eq:beta} to express the final result in terms of $\beta$ and find
\begin{align}
\label{eq:ppn-lightcurve}
\mu_\text{tot}(\beta) &= \frac{\beta^2 + (A_1/2)}{\beta\sqrt{\beta^2 + A_1}} + \mu_\text{tot,2} \, \epsilon^2 +  \mu_\text{tot,3} \, \epsilon^3 \, ,
\end{align}
where we have now derived the leading-order term for the microlensing magnification as mentioned in the Introduction, and the next-to-leading order term linear in $\epsilon$ vanishes identically. The expressions for $\mu_\text{tot,2}$ and $\mu_\text{tot,3}$ are a bit lengthy and displayed in appendix \ref{app:ppn}.

After the dust has settled, let us make a few comments on how this perturbative scheme will be used in what follows. Given the source position $\beta$, Eq.~\eqref{eq:beta} gives the leading-order positions for the lensed images as the two solutions of the resulting quadratic equation. In a typical lensing setup, the source position $\beta$ is described by the angular deviation of the source from the optical axis spanned by the observer and the lens. In the microlensing setting (which we will adapt later), however, the black hole lens is assumed to be moving, which implies that the effective angle $\beta$ will pick up that time dependence.

Schematically noting this time dependence as $\beta \mapsto \beta(T)$, we then define the light curve as the total magnification viewed as a function of time, that is,
\begin{align}
\mu_\text{tot}(T) = \mu_\text{tot}\left[\beta(T)\right] \, ,
\end{align}
where $T$ is a parameter that labels different lens positions. The quantity $|\beta(T)|$ is strictly positive, if and only if the lens never crosses the observer--star optical axis.

Let us now isolate the leading order impact of the regulator $\ell > 0$ on the magnification. We define the dimensionless quantity
\begin{align}
\xi = \frac{\ell}{GM} \, ,
\end{align}
which captures the deviations from general relativity (see Table \ref{table:ppn-coefficients} for the PPN expansion coefficients $a_i$ and $b_i$ of the non-singular metrics discussed in this work). We then expand  the difference between the total magnification \eqref{eq:ppn-lightcurve} for the Schwarzschild metric (denoted ``S'') and the Hayward metric, Minkowski core metric, and Simpson--Visser metric (denoted ``H,'' ``M,'' and ``W,'' respectively) in $\xi$ to leading order and find
\begin{align}
\label{eq:delta-mu-sh}
\delta\mu_\text{SH} &\equiv \mu_\text{tot,S}(\beta) - \mu_\text{tot,H}(\beta) \\
&= -\frac{15\pi \epsilon^3\xi^2}{16\beta} + \mathcal{O}\left( \xi^3 \right) \, , \nonumber \\
\label{eq:delta-mu-sm}
\delta\mu_\text{SM} &\equiv \mu_\text{tot,S}(\beta) - \mu_\text{tot,M}(\beta) \\ 
&= \left( -\frac{\epsilon^2 \xi}{6144\beta}\right) \Bigg[ \frac{196608}{(4+\beta^2)^{3/2}} - \frac{51840 \pi^2}{(4+\beta^2)^{5/2}} \nonumber \\
&\hspace{20pt} - \left( 38304 + 18432 d^2 - 6075 \pi^2 \right) \pi \epsilon \Bigg] + \mathcal{O}(\xi^2) \, , \nonumber \\
\label{eq:delta-mu-sw}
\delta\mu_\text{SW} &\equiv \mu_\text{tot,S}(\beta) - \mu_\text{tot,W}(\beta) \\
&= \left( \frac{\epsilon^2 \xi^2}{12288\beta}\right) \Bigg[ \frac{32768}{(4+\beta^2)^{3/2}} - \frac{17280 \pi^2}{(4+\beta^2)^{5/2}} \nonumber \\
&\hspace{20pt} - \left( 17568 + 6144 d^2 - 2025 \pi^2 \right) \pi \epsilon \Bigg] + \mathcal{O}(\xi^3) \, . \nonumber
\end{align}
Note that depending on the non-singular metric under consideration, the correction terms enter quadratic order in $\epsilon$ (for the cases M and W) or at cubic order in $\epsilon$ (for the well-known case H). Conversely, their dependence on the dimensionless regulator $\xi$ is linear (for case M) or quadratic (for the cases H and W).

Once the lensing geometry is known, the dimensionless source angle $\beta = \mathcal{B}/\vartheta_\text{E}$ can be inserted in order to estimate the quantitative result. As expected, these expressions are suppressed by powers of $\epsilon$ and $\xi$. However, we emphasize that they scale with $1/\beta$, which can amplify them for microlensing scenarios in which the source angle is small. In this case, our formulas \eqref{eq:delta-mu-sh}--\eqref{eq:delta-mu-sw} can be used to directly constrain the new physics parameter $\ell$ (given the prior of a non-singular metric). In terms of dimensionful parameters,
\begin{align}
\begin{split}
\frac{\epsilon^p}{\beta} &= \left[ \sqrt{ \frac{d_\text{ol} d_\text{os}}{4GM\,d_\text{ls}} } \arctan\left( \frac{GM}{d_\text{ol}} \right) \right]^p \frac{\vartheta_\text{E}}{\mathcal{B}} \\
&\approx \frac{1}{2^{p-1}} \sqrt{ \frac{ (GM)^{p+1} \, d_\text{os}^{p-1} }{ d_\text{ol}^{p+1} d_\text{ls}^{p-1} } } \, \frac{1}{\mathcal{B}} \, , \quad p = 2, 3 \, .
\end{split}
\end{align}

 If the above formulas are applied at the moment of smallest effective source angle, the result is independent of the black hole proper motion parameters, which further simplifies the considerations.

In contrast, other analyses of the perturbative expressions \eqref{eq:ppn-lightcurve}, such as the study of the full-width-half-maximum (FWHM), are exacerbated by the non-trivial $\beta$-dependence and are sensitive to the black hole proper motion parameters. For this reason, detailed studies involving astrometric parameters of the lens itself lie beyond the scope of this work.

\subsection{OGLE-2011-BLG-0462/MOA-2011-BLG-191}
\label{sec:exp}

\begin{figure}[!htb]
\centering
\includegraphics[width=0.48\textwidth]{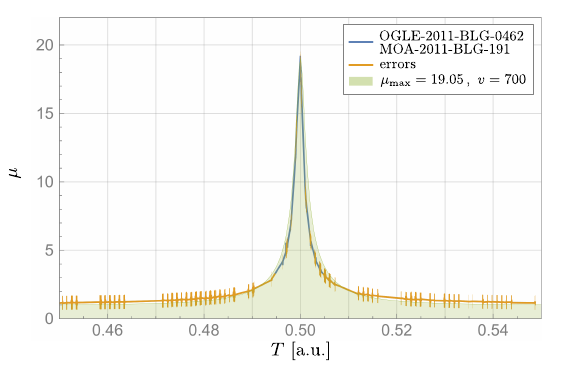}
\caption{The microlensing event OGLE-2011-BLG-0462 / MOA-2011-BLG-191, expressed in the conventions of this article. The errors are shown in orange, the curve between measured intensities is the solid line in blue, and the green area corresponds to the naive fit \eqref{eq:naive-fit}.}
\label{fig:microlensing-exp}
\end{figure}

In order to estimate the size of these effects, let us apply the developed formulae to a microlensing event. Recently, a stellar-mass rogue black hole has been detected via microlensing \cite{OGLE:2022gdj,Lam:2022vuq,Lam:2023}; the estimated parameters of this event are
\begin{align}
d_\text{ol} &= 1.58 \pm 0.18 \, \text{kpc} \, , \\
d_\text{os} &= 5.9 \pm 1.3 \, \text{kpc} \, , \\
M_\bullet &= 7.15 \pm 0.83 \, M_\odot \, ,
\end{align}
corresponding to a lensing event against a background star in the Milky Way bulge, emitting no X-rays consistent with an isolated black hole \cite{Mereghetti:2022qgz}. We emphasize that the mass of the object was extracted using the astrometric shift of the lensed image, which is not discussed in the present work. We extract the microlensing data from the public repository \cite{OGLE:repository} and convert the Heliocentric Julian Date to our dimensionless parameter $T$ via the transformation
\begin{align}
T = \frac{\text{HJD} - 2455260}{2 \times 503} \, , 
\end{align}
and convert from magnitudes $M_\text{OGLE}$ to luminosity magnification $\mu$ with a bolometric correction $BC$ via
\begin{align}
\mu = 10^{-0.4 (M_\text{OGLE} - BC)} \, , \quad BC = 16.4 \, .
\end{align}
The raw data (with error bars), together with a na\"ive fit of the form
\begin{align}
& \mu(T) = \frac{2+\beta^2}{\beta\sqrt{\beta^2+4}} \, , \quad
\beta = \sqrt{ \mu_\text{max}^{-2} + v (1 - 2 T)^2} \, , \nonumber \\
& \mu_\text{max} \approx 19.05 \, , \quad v \approx 700 \, , \label{eq:naive-fit}
\end{align}
can be seen in Fig.~\ref{fig:microlensing-exp}. 

We extract the minimal angle of $\beta_{T=0.5} \approx 0.05$ and insert this into the expressions \eqref{eq:delta-mu-sh}--\eqref{eq:delta-mu-sw}, yielding 
\begin{align}
\delta\mu_{SH} &= -1.9 \times 10^{-24} \, \xi^2 \, \, \mu_S \, , \\
\delta\mu_{SM} &= -5.4 \times 10^{-17} \, \xi \, \mu_S \, , \\
\delta\mu_{SW} &= -2.0 \times 10^{-17} \, \xi^2 \, \mu_S \, ,
\end{align}
where we normalized with respect to the third-order Schwarzschild magnification $\mu_S$. The minus signs in the above expressions mean that non-singular metrics feature a larger magnification. For this microlensing event the relative magnification increase is very small, even if $\xi \sim 1$, largely due to the small $\epsilon \sim 8.6 \times 10^{-9}$.

\subsection{PPN lightcurves}
\label{sec:ppn-results}

\begin{figure}[!htb]
\centering
\includegraphics[width=0.48\textwidth]{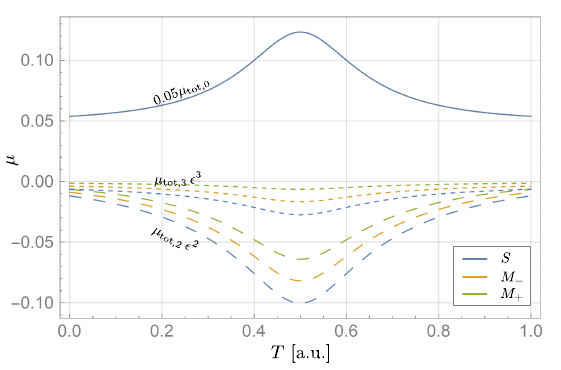}
\caption{The parametrized post-Newtonian (PPN) lightcurves (zeroth-order contributions: solid line; second-order contributions: dashed lines; third-order contributions: dotted lines) for the three cases of the Schwarzschild metric (S), the Minkowski core metric with a horizon ($M_-$) and without a horizon ($M_+$), for the parameters described in the text. Note that the zeroth-order curve is rescaled by a factor of $1/20$. Higher order PPN corrections reduce the magnification, but the presence of a regulator reduces that amount, leading to a net \emph{increase} in magnification.}
\label{fig:ppn-lightcurve}
\end{figure}

Having the PPN lightcurve at our disposal to fourth order \eqref{eq:ppn-lightcurve}, we can now track the influence of the correction terms stemming from the deviations from the Schwarzschild metric. For practical purposes, such correction terms show up highly suppressed, which is why we will only verify the general structure at this point. For illustrative purposes, we utilize here a microlensing setup corresponding to
\begin{align}
d_\text{ol} &= 50GM \, , \quad d_\text{ls} = 100GM \, , \\
x_\perp &= 20GM \, , \quad z_\perp = 5GM \, ;
\end{align}
this choice of parameters corresponds to the numerical values we will eventually choose in the second part of the paper. We can now plot the different order contributions $\mu_\text{tot,0}$, $\mu_\text{tot,2}\epsilon^2$, and $\mu_\text{tot,3}\epsilon^3$ for the Schwarzschild metric and the Minkowski core metric with $\ell/(2GM) = 0.7$ (subcritical case, with horizons) and $\ell/(2GM) = 1.3$ (the supercritical case, no horizon) for illustrative purposes; the result can be seen in Fig.~\ref{fig:ppn-lightcurve}.

The results are qualitatively similar between different non-singular models which is why we focus on the Minkowski core example. Since $a_1=b_1=1$ for all models considered, $\mu_\text{tot,0}$ agrees between all models since it only depends on $A_1 = 2(a_1+b_1)$. As expected, the next-to-leading-order contributions are heavily suppressed, but nevertheless distinct from the Schwarzschild case. Computing the relative magnification deviations, we find for this numerical setup
\begin{align}
\delta\mu_{SH} &= -0.02 \, \xi^2 \, \, \mu_S \, , \\
\delta\mu_{SM} &= -0.15 \, \xi \, \mu_S \, , \\
\delta\mu_{SW} &= -0.01 \, \xi^2 \, \mu_S \, ,
\end{align}
corresponding to percent-level increases in magnification. This again shows that the regulator $\ell$ increases magnification compared to the Schwarzschild case.

Last, we point out that for larger regulators the black hole horizon disappears, resulting in a horizonless geometry. Then, the PPN magnification setup is no longer applicable since the horizon scale is no longer of physical significance. We take this as another argument to develop a numerical setup towards microlensing.

\section{Ray tracing for static, spherically symmetric and asymptotically flat metrics}
\label{sec:ray-tracing}

With the perturbative machinery developed, let us now discuss a numerical approach. In static, spherically symmetric spacetimes \eqref{eq:metric}--\eqref{eq:asymptotically-flat} we may utilize the existence of a Killing vectors to introduce constants of motion. In asymptotically flat spacetimes we may utilize Schwarzschild coordinates $\{t,r,\theta,\varphi\}$, wherein the Killing vectors are given by
\begin{align}
\xi &= \partial_t \, \\
\rho_1 &= \partial_\varphi \, , \\
\rho_2 &= +\sin\varphi\partial_\theta + \cot\theta\cos\varphi \partial_\varphi \, , \\
\rho_3 &= -\cos\varphi\partial_\theta + \cot\theta\sin\varphi \partial_\varphi \, ,
\end{align}
resulting in the following constants of motion:
\begin{align}
E &\equiv (-1)g{}_{\mu\nu} u{}^\mu \xi{}^\nu = A\dot{t} \, , \\
L_1 &\equiv g{}{}_{\mu\nu} u{}^\mu \rho{}_1^\nu = C^2\sin^2\theta\dot{\varphi} \, , \\
L_2 &\equiv g{}_{\mu\nu} u{}^\mu \rho_2^\nu = +C^2 \dot{\theta}\sin\varphi + L_1 \cot\theta\cos\varphi \, , \\
L_3 &\equiv g{}_{\mu\nu} u{}^\mu \rho_3^\nu = -C^2 \dot{\theta}\cos\varphi + L_1 \cot\theta\sin\varphi \, .
\end{align}
An additional constant of motion is of course
\begin{align}
g{}_{\mu\nu} u{}^\mu u{}^\nu = 0 \, .
\end{align}
It is hence clear that this system is integrable, but for numerical evaluation not all equivalent formulations are equally efficient. Since we are only interested in the spatial aspects of the problem, we use the affine freedom of null geodesics and rescale their affine parameter such that $E=1$ and subsequently eliminate the equation of motion for $t(\lambda)$. It is possible to work with first-order differential equations via effective potentials, as done in the pioneering work by Luminet \cite{Luminet:1979nyg} for the imaging of accretion disks and that of Rauch and Blandford \cite{Rauch:1994} in the context of gravitational lensing of Kerr black holes. Working with \textsc{Mathematica}, we find that the utilization of $L_1$ to render the equation of motion for $\varphi(\lambda)$ first order leads to significant speedup, whereas implementing $L_2$ and $L_3$ (to obtain first-order equations for $r(\lambda)$ and $\theta(\lambda)$) actually slows the code down, most likely due to the non-linear nature of the radial equation as well as the more involved appearance of trigonometric functions in the polar equation. Moreover, the possible existence of multiple turning points in $r$ make it inconvenient to rephrase the radial differential equation in terms of an effective potential without previously identifying the location of turning points, which would have to be performed in a distinct way for each metric considered in this work. For this reason we opt to utilize the second-order approach in the radial and polar sector.

Last, given a local radiation density $n = n(r,\theta,\varphi)$, we define a simple radiative transport equation for the intensity $I(\lambda)$, and we hence arrive at the following set of first-order differential equations for $r(\lambda)$ and first-order differential equations for $\theta(\lambda)$, $\varphi(\lambda)$ and $I(\lambda)$:
\begin{equation} \left\{ \begin{aligned}
\ddot{r} &= \frac{B'}{2B} \dot{r}^2 - \frac{B}{2}\left[ \frac{A'}{A^2} - 2CC' \dot{\theta}^2 - \frac{2C'L_1^2}{C^3\sin^2\theta} \right] \, , \\
\ddot{\theta} &= - \frac{2C'\dot{\theta}\dot{r}}{C} + \cos\theta \frac{L_1^2}{C^4\sin^3\theta} \, , \\
\dot{\varphi} &= \frac{L_1}{C^2\sin^2\theta} \, , \\
\dot{I} &= n(r, \theta, \varphi) \, ,
\end{aligned} \right. \end{equation}

In order to accommodate arbitrary, three-dimensional densities for the radiative transport equation we chose not to eliminate $\theta$ entirely. While allowed due to the spherical symmetry of the black hole geometries under consideration in this paper, it would still necessitate the computation of the full three-dimensional trajectory due to the explicit dependence on all three Cartesian coordinates $x$, $y$, and $z$ in the radiation density. The effective polar angle would still be different for each ray-traced pixel (since it is determined by the plane of the black hole, the observer, and the pixel) which would lead to more numerical overhead. In practice, a resolution of $200 \times 200$ pixel is sufficient to generate images of sufficient accuracy, and we have verified that images of a larger resolution of $1000 \times 1000$ only provide marginal improvement. Since the primary goal of the numerical aspect of this work is to generate ray traced lightcurves, an enhanced angular resolution to capture faint details is not required. Moreover, no large-scale data analysis is performed, and for this reason the full, three-dimensional treatment is retained. Last, this will also streamline the future implementation of axially symmetric metrics with angular momentum in a straightforward manner.

In what follows, we will set up the appropriate boundary conditions for the numerical study of gravitational lensing in such a scenario. 

\subsection{Implementing the lensing setup}

\begin{figure}[!htb]
\centering
\includegraphics[width=0.48\textwidth]{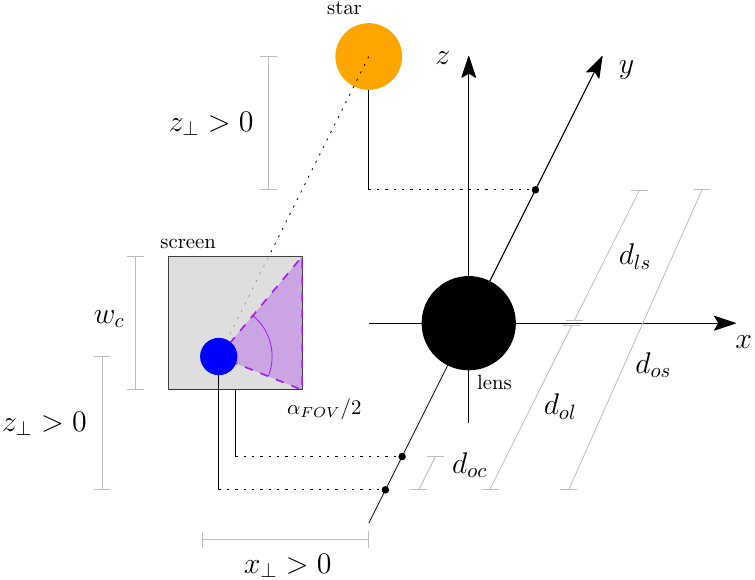}
\caption{Visualization of the ray tracing setup. Instead of a moving lens, it is computationally easier to model the lens in the origin and consider instead a moving star and screen that share the same $x$-coordinate for their center point. For the purposes of this paper we restrict the motion of the lens to be purely along the $x$-axis.}
\label{fig:ray-tracing}
\end{figure}

Since we are interested in ray tracing an image of the black hole, we formulate our initial value problem accordingly; see Fig.~\ref{fig:ray-tracing} for our geometric setup. The physically relevant scenario involves a  black hole that traverses the line-of-sight between a stationary star and the observer. For computational simplicity, we instead consider a moving line-of-sight between the star and the observer and keep the black hole at the coordinate origin and describe the black hole proper motion via a simple coordinate shift with an auxiliary time coordinate $T \in [0,1]$. We work in Cartesian coordinates $\{x, y, z\}$ and parametrize the black hole proper motion via
\begin{align}
\vec{x}_\bullet(T) = \big\{ -x_\perp(1-2T), 0, z_\perp \big\} \, ,
\end{align}
where $x_\perp > 0$ labels the initial horizontal position of the black hole. Note that $T$ merely serves as a label and is not a time parameter in the physical sense. The detector and the star are located at
\begin{align}
\vec{x}_\text{o}(T) &= (0, -d_\text{ol}, 0) - \vec{x}_\bullet(T) \, , \\
\vec{x}_\star(T) &= (0, d_\text{ls}, 0) - \vec{x}_\bullet(T) \, ,
\end{align}
respectively, where $d_\text{ol} > 0$ is the distance between the observer plane and the lens plane, and $d_\text{ls}$ is the distance between the lens plane and the source plane, see Fig.~\ref{fig:lensing}.

It is important to realize that for every $T$ the lensing setup changes. This is because the optical axis is formed by the line of sight between the observer and the lens, and due to the proper motion of the lens the distances to the lens are a function of time and need to be obtained from the bare parameters $d_\text{ol}$, $d_\text{ls}$, and $d_\text{os}$ via suitable projections onto the optical axis. For notational brevity we define the connection vectors
\begin{align}
\vec{x}_{\text{o}\bullet} \equiv \vec{x}_\bullet - \vec{x}_\text{o} \, , \quad
\vec{x}_{\text{o}\star} \equiv \vec{x}_\star - \vec{x}_\text{o} \, .
\end{align}
We then define the effective quantities
\begin{align}
d_\text{ol,\,eff}(T) &= |\vec{x}_{\text{o}\bullet}| \, , \\
d_\text{os,\,eff}(T) &= \frac{\vec{x}_{\text{o}\star}\cdot \vec{x}_{\text{o}\bullet}}{|\vec{x}_{\text{o}\bullet}|} \, , \\
d_\text{ls,\,eff}(T) &= d_\text{os,\,eff}(T) - d_\text{ol,\,eff}(T) \, , \\
\vartheta_\text{E,\,eff}(T) &= \sqrt{ \frac{4GM d_\text{ls,\,eff}}{d_\text{ol,\,eff} \, d_\text{os,\,eff} } } \, , \\
\beta(T) &= \frac{1}{\vartheta_\text{E,\,eff}} \arccos \left( \frac{\vec{x}_{\text{o}\bullet}\cdot\vec{x}_{\text{o}\star}}{|\vec{x}_{\text{o}\bullet}| |\vec{x}_{\text{o}\star}|} \right) \, ,
\end{align}
where we have suppressed the $T$-dependence on $\vec{x}_\bullet$, $\vec{x}_\text{o}$, $\vec{x}_\star$ and $\vartheta_\text{E,\,eff}$ for brevity, and the physical angle is obtained after a rescaling with the Einstein angle, $\mathcal{B} = \beta \vartheta_\text{E}$. In particular, for $z_\perp \not= 0$ one has $\beta > 0$ for all times $T$, meaning that the lens does not intersect the optical axis.

In what follows, we will first set $d_\text{ol}$ and $d_\text{ls}$ as free parameters, and compute $d_\text{os} = d_\text{ol} + d_\text{ls}$, which gives $\vartheta_\text{E}$. The Einstein angle is helpful  to adjust the field of view $\alpha_\text{FOV}$ to a convenient value, which is typically several $\vartheta_\text{E}$. Next, we can fix the arbitrary distance of the detector plane to the observer plane ($d_\text{oc} = GM$ in our code) to determine the physical width and height of the quadratic screen, $w_\text{c} = 2d_\text{oc}\tan(\alpha_\text{FOV}/2)$. The proper motion of the lens can be chosen freely, but care has to be taken that for early and late times $x_0$ is not outside the field of view. Each ray tracing setup can hence be determined by the following set of numbers:
\begin{align}
\big\{ \alpha_\text{FOV}, d_\text{ol}, d_\text{ls}, x_0, z_0, T \big\}
\end{align}
Moreover, we subdivide the screen into pixels with a resolution of $\rho$ pixels (for microlensing, $\rho=256$ is sufficient, but for more detailed imaging we instead utilize $\rho=1024$). Each ray is cast from the observation point aiming for a single pixel, and the integrated intensity is then reported back numerically as the brightness information for that pixel. Since gravitational bending of light has no chromatic aberration, we do not consider any frequency dependence and model the intensity function $n$ phenomenologically (more on that later). The brightness levels are not normalized \textit{per se}, so we will either normalize to the brightest pixel in each image (for single images), or to an otherwise defined brightness level across a sequence of pixels (e.g. the total brightness of the star in the absence of the lens, and the otherwise identical distance).

On the computational side, since ray tracing is a parallelizable algorithm, we make extensive use of \textsc{Mathematica}'s built-in ParallelTable command, such that on a desktop computer with 24 cores a $\rho=1024$ picture takes about 20 minutes to render. We implemented a recursive algorithm featuring a simple adaptive-mesh ray tracing, which should theoretically be faster for images with sharply concentrated illuminated areas, but found that the recursive structure did not lend itself well to parallelization, resulting in a performance drop.

We check for collisions with the black hole horizon by $B(r(\lambda)) < 0.05$ (varying this value has little effect on the images), and discard all geodesics that exceed an affine parameter of $\lambda > \lambda_\text{max} = 2d_\text{os}$ or reach $r > r_\text{max} = 2d_\text{os}$, whichever happens sooner, where again variations of this cutoff lead to negligible changes in the resulting images. As additional quality-of-life features we also directly project the shape of the black hole event horizon and the Einstein radius onto the screen. For this perspective mapping we utilize simple Euclidean geometry, since they are only included as reference surfaces for convenience---more on that in the next sections.

\section{Test: Strong lensing}
\label{sec:strong}

Let us now test the developed ray tracing machinery in the case of strong lensing as applied to stars and accretion disks. We will only record a single image and therefore fix the value of $T$ in each scenario.

\subsection{Lensing of a star}
\begin{figure*}[!htb]
\centering
\includegraphics[scale=1,valign=c]{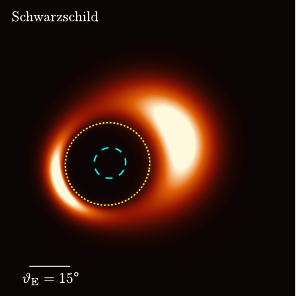} \quad
\includegraphics[scale=1,valign=c]{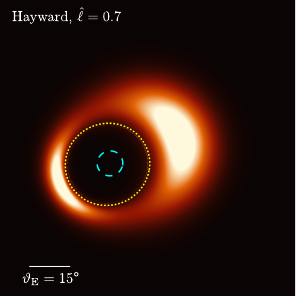} \quad
\includegraphics[scale=1,valign=c]{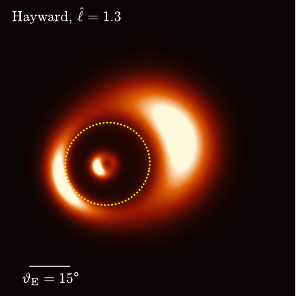} \quad
\includegraphics[scale=0.163,valign=c]{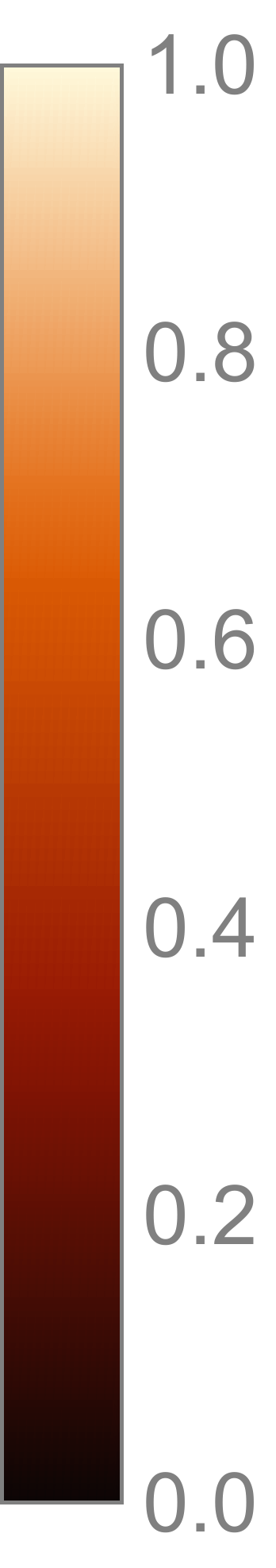}
\caption{Luminosity of a lensed star under passage of a black hole described by the scenarios $S$, $H_-$ and $H_+$ at $T=0.25$. The Einstein angle is visualized at the bottom left, the dashed cyan line is the projected outline of the black hole horizon (if present), and the yellow dashed line is the projected Einstein radius. The images are individually normalized to their brightest spot and then amplified by a factor of 2.5 to enhance contrast. The subcritical Hayward image features a smaller black hole horizon, and novel features appear in the center of the horizonless supercritical Hayward case on the rightmost panel.}
\label{fig:star-test}
\end{figure*}

We model the radiation density of a star of radius $r_\star$ as constant in its interior and exponentially suppressed in its exterior,
\begin{align}
\label{eq:density-star}
n(r,\theta,\varphi) = n_0 \begin{cases}
1 \, , \quad & r \leq r_\star \, , \\[5pt]
\displaystyle \exp\left[ - \frac{(r_\text{s}-r_\star)^2}{\omega^2} \right] \, , & r > r_\star \, .
\end{cases}
\end{align}
In the above, we defined the distance from the center of the star according to
\begin{align}
\begin{split}
r_\text{s} = r_\text{s}(\lambda, T) &= \bigg\{ [x(\lambda) - x_\star(T)]^2 \\
			 &\hspace{25pt} + [y(\lambda) - y_\star(T)]^2 \\
			 &\hspace{25pt} + [z(\lambda) - z_\star(T)]^2 \bigg\}^{1/2} \, ,
\end{split}
\end{align}
with the Euclidean distances
\begin{align}
x(\lambda) &= r(\lambda)\sin\theta(\lambda)\cos\varphi(\lambda) \, , \\
y(\lambda) &= r(\lambda)\sin\theta(\lambda)\sin\varphi(\lambda) \, , \\
z(\lambda) &= r(\lambda)\cos\theta(\lambda) \, ,
\end{align}
For simplicity we will choose $r_\star = 3GM$ and $\omega = 5 GM$. This setup can now be validated by fixing the parameter $T$, and the result is displayed in Fig.~\ref{fig:star-test} for the Schwarzschild metric and the overcritical Hayward metric, with superimposed black hole shape and Einstein radius.

\subsection{Accretion disks}
Similar to the previously discussed case we will again utilize a simple, phenomenologically motivated model for an accretion disk as found, for example, in Ref.~\cite{Eichhorn:2021iwq,Eichhorn:2022bbn}, taking the form
\begin{align}
\begin{split}
n(r,\theta) &= n_0 \left(\frac{GM}{r}\right)^\alpha \exp \left[ -\frac{\cos^2\theta}{2h^2} \right] \\[5pt]
&\hspace{11pt} \times \begin{cases} \displaystyle \exp\left[ - \frac{(r-r_\text{cut})^2}{\omega^2} \right] \, , \quad & r \leq r_\text{cut} \, , \\[5pt]
1 \, , & r > r_\text{cut} \, .
\end{cases}
\end{split}
\end{align}
We choose the parameters $\alpha = 1.5$, $r_\text{cut} = 3GM$, $\omega = 1/\sqrt{12}GM$, and $h=0.2$, which are inspired by the choices in Ref.~\cite{Eichhorn:2022bbn}; the results can be seen in Fig.~\ref{fig:disk-test} for the Schwarzschild metric for two different observer inclinations. See the appendix for an exhaustive comparison.

\begin{figure*}[!htb]
\centering
\includegraphics[scale=1,valign=c]{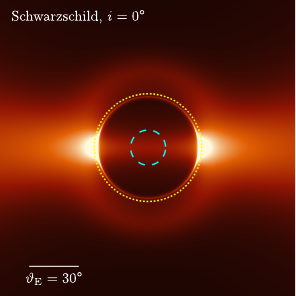} \quad
\includegraphics[scale=1,valign=c]{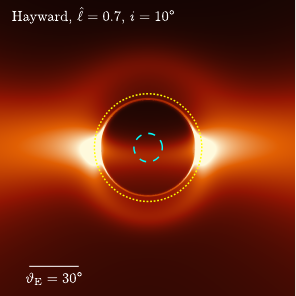} \quad
\includegraphics[scale=1,valign=c]{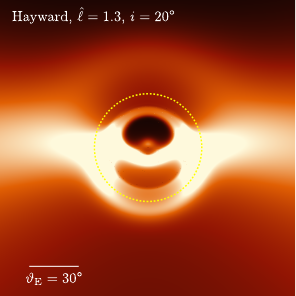} \quad
\includegraphics[scale=0.163,valign=c]{legend.pdf}
\caption{Accretion disk luminosity (in arbitrary units) for the scenarios $S$, $H_-$ and $H_+$ at different inclinations ($i=0^\circ$ corresponds to an edge-on view, and $i>0$ describes the angle out of the equatorial plane). The Einstein angle is visualized at the bottom left, the dashed cyan line is the projected outline of the black hole horizon (if present), and the yellow dashed line is the projected Einstein radius. The images are individually normalized to their brightest spot and then amplified by a factor of 2.5 to enhance contrast. The subcritical Hayward image features a smaller black hole horizon, and novel features appear in the horizonless supercritical Hayward case on the right.}
\label{fig:disk-test}
\end{figure*}

\section{Microlensing}
\label{sec:microlensing}

We proceed as follows: after fixing the lensing geometry we first render the background star without the influence of the black hole by taking the $GM \rightarrow 0$ limit of the geodesic equation, thereby generating a reference image of which the luminosity is determined via simple summation over all pixels. Subsequently, we vary $T \in [0,1]$ and record single images for each time step, and determine the relative intensity by normalizing the pixel luminosities with respect to the initial reference image. We call the result of this procedure the finite-source lightcurve, and will describe its properties in what follows. The precise values utilized for all lightcurves described in this paper are summarized in Table~\ref{table:numerical-parameters} in appendix \ref{app:images} along with a subset of images in Fig.~\ref{fig:lightcurve-overview}.

\subsection{Validation}

\begin{figure}[!h]
\centering
\includegraphics[width=0.48\textwidth]{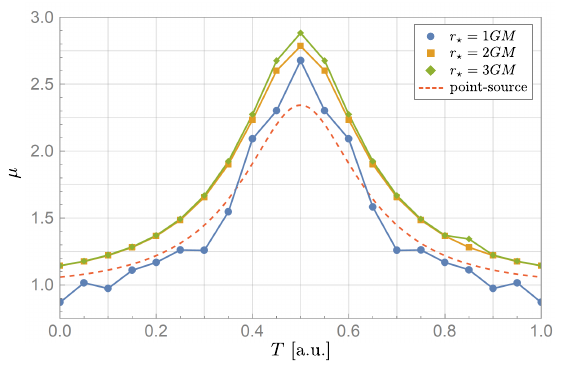}
\caption{We plot the numerical lightcurves for the Schwarzschild metric for identical lensing systems but varying star size. As the star reduces in size, the numerical data approaches the theoretical point source prediction (dashed line), as expected. However, numerical uncertainties increase as the star radius shrinks, necessitating a higher rendering resolution. In the above, we chose a resolution of $256\times256$.}
\label{fig:validation}
\end{figure}

Initially, before discussing non-singular objects, we perform a simple run for the Schwarzschild geometry and for simplicity we fix $d_\text{ol} = 100GM$ and $d_\text{ls} = 200GM$ and set $\alpha_\text{FOV} = 6\vartheta_\text{E}$. We initially record the lightcurve for the star radius at $r_\star = 3 GM$ and the parameter $\omega=5GM$ (which describes the falloff of the star luminosity), and then consider $r_\star = 2GM$ and $\omega=(10/3) GM$, and finally $r_\star = GM$ and $\omega = (5/3)GM$ (keeping the ratio $r_\star/\omega = 5/3$ constant in order to assure a consistent shape of the star geometry). Comparing the resulting lightcurves to the point-source prediction obtained in the PPN context in Sec.~\ref{sec:ppn-results}, we find the following (see Fig.~\ref{fig:validation} for the lightcurves, and appendix \ref{app:images} for the individual images):
\begin{itemize}
\item As the star radius decreases, the numerical lightcurves approach the point-source PPN lightcurve.
\item At the same time, the numerical results become less reliable at smaller star size, requiring a larger resolution.
\end{itemize}
Both of these behaviors are expected, and we take this as a validation of our numerical procedure. The complete set of parameters for these lightcurves are listed in Table~\ref{table:numerical-parameters} under $V_1$, $V_2$, and $V_3$. 

\subsection{Near zone}

\begin{figure}[!t]
\centering
\includegraphics[width=0.48\textwidth]{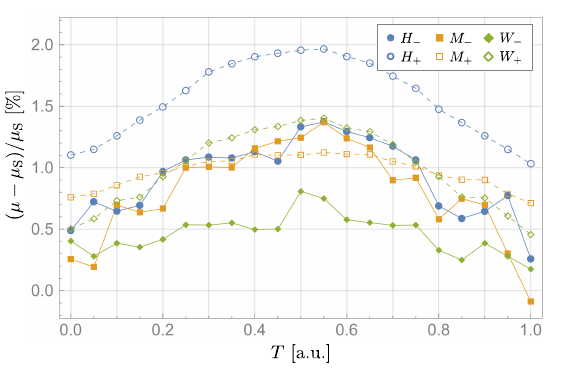}
\caption{Microlensing lightcurves in the near-zone scenario for the Hayward metric $H_\pm$, Minkowski core metric $M_\pm$, and the Simpson--Visser metric $W_\pm$. The subscript ``+'' denotes the horizonless case, and the subscript ``-'' denotes the non-singular black hole case. Generally, the horizonless cases exhibit a stronger magnification by approximately $1\%$, while the difference between the models themselves is around $0.5\%$.}
\label{fig:lightcurves-numerics-near}
\end{figure}

Let us now fix a set of values that harnesses the finite-size capability of our code, $d_\text{ol} = 20GM$ and $d_\text{ls} = 10GM$ (see Table~\ref{table:numerical-parameters} for the dataset $N_1\dots N_7$). The resulting lightcurves differ appreciably for $T \sim 0.5$, with the horizonless cases featuring central, bright spots related to the absence of a horizon as could already have been anticipated from Fig.~\ref{fig:star-test}. Numerics are rather stable at these close distances, but instead of plotting the microlensing lightcurves directly, it is instructive to consider instead the relative magnification
\begin{align}
\mu_\text{rel} \equiv \frac{\mu - \mu_\text{S}}{\mu_\text{S}} \, ,
\end{align}
where $\mu_\text{S}$ denotes the Schwarzschild magnification; see the resulting relative magnifications in Fig.~\ref{fig:lightcurves-numerics-near}.

\subsection{Far zone}

A more realistic setup is considered in this far-zone scenario, where we pick $d_\text{ol} = 50GM$ and $d_\text{ls} = 100GM$ (see Table~\ref{table:numerical-parameters} for the dataset $F_1\dots F_7$). While the general features can still be resolved in similarity to the near-zone case, the strength of the deviations is suppressed. Moreover, numerical stability becomes more challenging. To see this, consider for example the time $T=0.05$ in the $H_-$ model, where we record a magnification of $\mu=1.241$. However, this value differs by more than $5\%$ from its neighboring timestamps at $T=0.00$ and $T=0.10$ (around $\mu=1.178$). While there are no convergence issues with the code, the utilized resolution of $256\times 256$ makes the recorded brightness susceptible to aliasing effects close to the boundary of the glowing centroid and the surrounding darkness. Even with a few pixels being affected, the stark brightness difference can quickly amount several percent of systematic error.

\begin{figure}[!htb]
\centering
\includegraphics[width=0.48\textwidth]{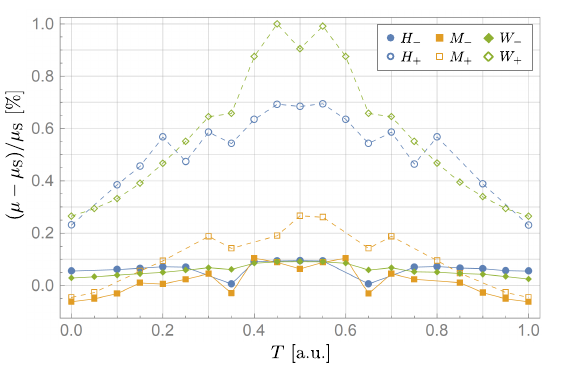}
\caption{Microlensing lightcurves in the far-zone scenario for the Hayward metric $H_\pm$, Minkowski core metric $M_\pm$, and the Simpson--Visser metric $W_\pm$. The subscript ``+'' denotes the horizonless case, and the subscript ``-'' denotes the non-singular black hole case. Similarly to the near-zone scenario, the horizonless cases exhibit a stronger magnification while the difference between the models themselves is smaller.}
\label{fig:lightcurves-numerics-far}
\end{figure}

We re-ran the questionable data point at $T=0.05$ in the dataset $F_2$ at doubled resolution of $512 \times 512$ and recorded a new normalization value without the black hole, resulting in a magnification value of $\mu=1.179$ that is entirely consistent with its neighboring values at $T=0.00$ and $T=0.10$. Since a simple doubling of resolution quadruples the running time, we opted against re-running the entire datasets at a higher resolution and instead manually removed the flawed datapoints from the dataset. In total, more than 15 datapoints were affected. With this numerical issue addressed, we can now plot the different lightcurves, see Fig.~\ref{fig:lightcurves-numerics-far}.

\section{Summary and conclusions}
\label{sec:conclusion}

This paper describes, for the first time, the microlensing of non-singular black holes (and their compact, horizonless counterparts) for finite-size sources. After extending the work by Keeton and Petters in the parametrized post-Newtonian framework to fourth order, we demonstrated that our numerical approach reproduces the theoretical predictions as the size of the source shrinks, as expected. Taking this as a validation of our numerical setup, we then explored the resulting finite-source microlensing lightcurves for various non-singular black hole models in the near-zone and far-zone regime. We find throughout that the existence of a regulator length scale $\ell > 0$ tends to \emph{increase} the magnification recorded in finite-source lightcurves, similar to the point-source predictions.

This is consistent with theoretical expectations:\footnote{We thank our Referee for bringing this to our attention.} Within the description of gravitational lensing via Fermat's principle, the image magnification is proportional to the inverse curvature of the local time-delay surface (see e.g.~\cite{Narayan:1996ba} for a pedagogical reference as well as Refs.~\cite{Schneider:1985,Blandford:1986,Nityananda:1992}). Since that curvature is smaller for smoothed-out matter distributions (such as the non-singular metrics considered in the present paper), this would then lead to a larger magnification.

Serving as a proof of principle, this work is far from complete. Numerical investigations of the far zone have shown the need for enhanced resolution. While a resolution of $1024 \times 1024$ can be obtained by a 24-core desktop computer in around 20 minutes, finer time steps and even larger resolutions are still computationally intensive, necessitating the migration of the numerical ray tracer into high-performance computing architecture. Written entirely in \textsc{Mathematica}, however, this is relatively straightforward since the performance of ray tracing individual geodesics is a highly parallelizable task.

In a wider context, we would also like to point out that the presented approach can be relatively easily extended to describe the related physical lensing processes of \emph{femtolensing} and \emph{picolensing}, which is what we want to briefly demonstrate in the following sections.

\subsection{Femtolensing}

If the source is modelled in a time-dependent manner via a simple generalization of Eq.~\eqref{eq:density-star} such that
\begin{align}
n(t,r,\theta,\varphi) = e^{-(t-t_0)^2/\tau^2} n(r,\theta,\varphi) \, ,
\end{align}
we may think of the above as a crude approximation of a gamma-ray burst. Assuming coherent emission of radiation over the characteristic time scale $\tau$ we may then endow each light ray with a phase, which can numerically be achieved by integrating the time variable,
\begin{equation} \left\{ \begin{aligned}
\ddot{r} &= \frac{B'}{2B} \dot{r}^2 - \frac{B}{2}\left[ \frac{A'}{A^2} - 2CC' \dot{\theta}^2 - \frac{2C'L_1^2}{C^3\sin^2\theta} \right] \, , \\
\ddot{\theta} &= - \frac{2C'\dot{\theta}\dot{r}}{C} + \cos\theta \frac{L_1^2}{C^4\sin^3\theta} \, , \\
\dot{t} &= \frac{1}{A} \, , \\
\dot{\varphi} &= \frac{L_1}{C^2\sin^2\theta} \, , \\
\dot{I} &= n(t,r, \theta, \varphi) \, ,
\end{aligned} \right. \end{equation}
where we still set $E=1$. Assuming now further that the wavelength of the radiation is much smaller than that of the astrophysical lens (which is guaranteed for keV radiation stemming from gamma-ray bursts), we can assign to each light ray a complex phase
\begin{align}
\Phi = e^{i \Omega t_\text{arrival} } \, ,
\end{align}
where $\Omega$ depends is the angular frequency of the arriving radiation, and $t_\text{arrival}$ is a suitably defined arrival time,
\begin{align}
t_\text{arrival} \approx d_\text{ol} + d_\text{ls} \, .
\end{align}
Each pixel thereby contributes to the intensity of the final image via interference due to the phase $\Phi$. Femtolensing, unlike microlensing, does not require measurement over long periods of time---rather, it uses the fact that gamma-ray bursts usually emit a broad range of wavelength during an emission. Since the interference pattern depends on the frequency of the radiation, by tracking the received intensity as a function of frequency one arrives at a characteristic intensity-frequency relation which can be compared with observed gamma-ray spectra \cite{Gould:1992}.

\begin{figure*}[!htb]
\centering
\includegraphics[width=0.27\textwidth]{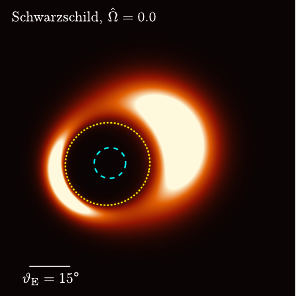} \quad
\includegraphics[width=0.27\textwidth]{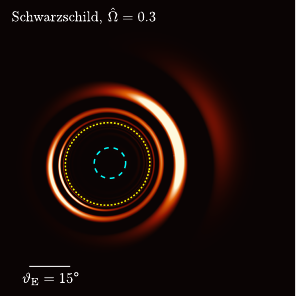} \quad
\includegraphics[height=0.27\textwidth]{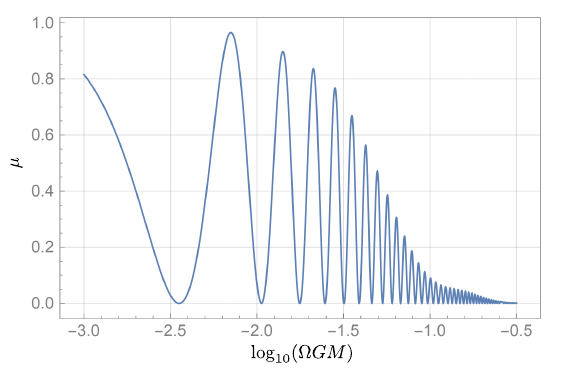}
\caption{Using ray tracing methods, we visualize the relative phases of a lensed gamma-ray source for two given dimensionless frequencies (left and center panel, $\hat{\Omega} \equiv \Omega GM$). In the right panel, we plot the total magnification as a function of frequency, which exhibits typical interference patterns.}
\label{fig:femto-test}
\end{figure*}

In order to visualize this, we work again in the near-zone regime and extend our numerical algorithm to integrate the time of arrival as well. Doing so, we can then visualize the incoming phase by multiplying the rendered image with $\cos^2(\Omega\Phi t)$, where $t_0$ is the moment of gamma-ray-burst eruption, showing that several parts of the image have opposite phase if the frequency $\Omega$ is small enough: see Fig.~\ref{fig:femto-test} for an example. While clearly not an exhaustive study, femtolensing remains a possible avenue to search for otherwise hard-to-probe dark matter regimes \cite{Barnacka:2012bm}, but the experimental feasibility of this approach has recently been questioned \cite{Katz:2018zrn}. It would certainly be interesting to investigate femtolensing and its sensitivity to probe for non-singular black holes further with the methods presented in this work.

\subsection{Picolensing}

\begin{figure}[!htb]
\centering
\includegraphics[width=0.42\textwidth]{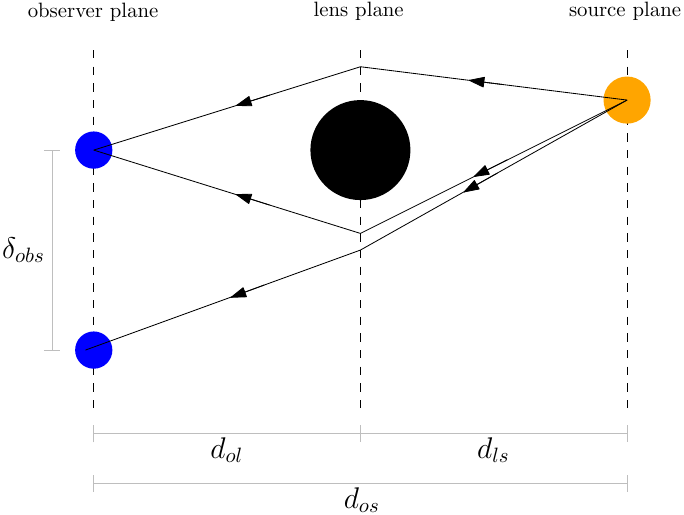}
\caption{We sketch a typical picolensing setup: two observers record the luminosity of a distant source. If a gravitational lens is present in between the observers and the source, the perceived brightness will vary strongly with their relative distance $\delta_\text{obs}$, assuming the observers are positioned in such a way that one observer is within the Einstein ring radius and the other one outside of it.}
\label{fig:sketch-picolensing}
\end{figure}

\begin{figure}[!h]
\centering
\includegraphics[width=0.48\textwidth]{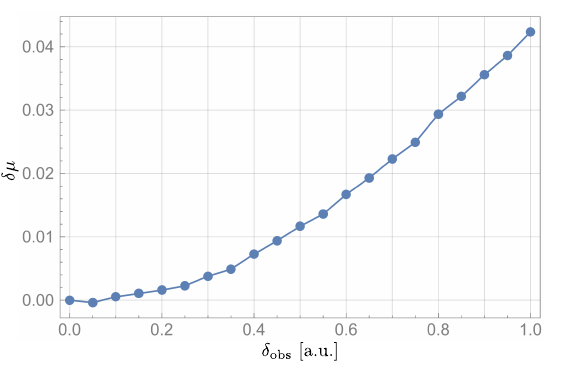}
\caption{We plot the relative magnification of a luminous background source as measured by two distant observers in the same instant of time, but as a function of the observer separation $\delta_\text{obs}$. The numerical simulation shows that it exhibits a dependence on the observer separation as expected from the picolensing effect.}
\label{fig:plot-picolensing}
\end{figure}

Picolensing utilizes the fact that for a short gamma-ray burst, two distant observers may record starkly different luminosities if one of the observers is located close to a lensing maximum of an dark, passing object \cite{Kolb:1995bu}---see Fig.~\ref{fig:sketch-picolensing}. It has recently been argued that picolensing is sensitive to probe for primordial black hole dark matter in the asteroid window $2 \times 10^{-16} M_\odot < M < 5 \times 10^{-12} M_\odot$ \cite{Fedderke:2024wpy}. Denoting the measured magnifications of two observers as $\mu_1$ and $\mu_2$, the relative magnification,
\begin{align}
\delta\mu = \frac{\mu_1-\mu_2}{\mu_1+\mu_2} \, ,
\end{align}
will depend on their separation $\delta_\text{obs}$. Such a scenario can be implemented in our scenario by first fixing the black hole position, and then moving the position of the star and observer separately---see Fig.~\ref{fig:plot-picolensing} for a visualization of the observer distance-dependent perceived magnification.

\subsection{Further work}

The presented method reproduces lightcurves reliably at intermediate distances, and tracks the impact of deviations from general relativity. It would be desirable to introduce a notion of numerical error to the relative magnification, as became obvious from the comparative study of the near-zone and far-zone regime. Other further improvements include the addition of angular momentum, which we expect will not affect performance significantly if we restrict ourselves to rotating non-singular geometries that feature a Carter constant (which is the case for many of them); these results could then supplement existing shadow computations \cite{Younsi:2016azx}. A more realistic modelling of the lens trajectory is also desirable, deviating from the straight-path trajectory utilized in this study. By implementing an algorithm that determines the centroid of the lensed image, the gravitational lensing parallax could be used to extract the lens mass numerically. While these topics are of physical interest, there is also a logistical opportunity. Namely, the data infrastructure of this project could be improved by converting the lightcurves data from proprietary CSV files (as used currently) into the community-documented formats of FITS or ECSV (the latter being useful for straightforward integration with the \textsc{astropy} Python library. Moreover, realistic conversions from relative magnification into astrophysical magnitudes seem desirable.

Black hole dark matter \cite{Carr:2020xqk,Green:2020jor} is constrained by lensing \cite{Carr:2020gox,Zumalacarregui:2017qqd,Garcia-Bellido:2017imq,Niikura:2017zjd,CrispimRomao:2024nbr} and extragalactic gamma ray flux \cite{Calza:2024fzo,Asmanoglu:2025agc}, but to date the asteroid mass window remains open \cite{Gorton:2024cdm}. As new data is expected from Subaru/HSC and the Vera C.~Rubin Observatory, we plan to extend our ray tracing code to contribute towards further probing that open window in the context of non-singular black hole models.

\section{Acknowledgements}
JB is grateful for support as a Fellow of the Young Investigator Group Preparation Program, funded jointly via the University of Excellence strategic fund at the Karlsruhe Institute of Technology (administered by the federal government of Germany) and the Ministry of Science, Research and Arts of Baden-W\"urttemberg (Germany).

\pagebreak

\appendix

\begin{widetext}

\section{Parametrized post-Newtonian framework at fourth order}
\label{app:ppn}

The perturbative parameter $\epsilon \ll 1$ is defined as
\begin{align}
\epsilon = \frac{\vartheta_\bullet}{\vartheta_\text{E}} = \sqrt{ \frac{d_\text{ol} d_\text{os}}{4GM\,d_\text{ls}} } \arctan\left( \frac{GM}{d_\text{ol}} \right) ~ \ll 1 \, .
\end{align}
The expansion for the image angle $\theta$ then takes the form
\begin{align}
\theta = \theta_0 + \theta_1 \epsilon + \theta_2 \epsilon^2 + \theta_3 \epsilon^3 + \mathcal{O}(\epsilon^4) \, ,
\end{align}
where the coefficients are
\begin{align}
\beta &= \theta_0 - \frac{A_1}{4\theta_0} \, , \\
\theta_1 &= \frac{A_2}{A_1 + 4\theta_0^2} \, , \\
\theta_2 &= \frac{1}{3\theta_0(A_1 + 4\theta_0^2)} \bigg( A_1^3 + 3 A_3 - A_1^3 d^2  + 12 A_1^2 (d-1) d \theta_0^2 + 8 A_1 d^2 \theta_0^4 + \frac{3 A_1 A_2^2}{(A_1 + 4 \theta_0^2)^2}  - \frac{ 6 A_2^2}{A_1 + 4 \theta_0^2} \bigg) \, , \\
\theta_3 &=  \frac{1}{3\theta_0^2(A_1 + 4\theta_0^2)} \bigg\{ \hspace{10pt} 3 A_1^2 (2 A_2^3 - 3 A_1 A_2 A_3 + A_1^2 A_4) \\
&\hspace{85pt} + 4 A_1 [21 A_2^3 - 33 A_1 A_2 A_3 + 12 A_1^2 A_4 + A_1^4 A_2 (d-1) (2d-1)] \theta_0^2 \nonumber \\
&\hspace{85pt} + 8 \{42 A_2^3 - 78 A_1 A_2 A_3 + 36 A_1^2 A_4 +  A_1^4 A_2 [10 + DD (11d-18)]\} \theta_0^4 \nonumber \\
&\hspace{85pt} + 32 \{-30 A_2 A_3 + 24 A_1 A_4 + A_1^3 A_2 [14 + 5 d (5d - 6)]\} \theta_0^6 \nonumber \\
&\hspace{85pt} + 128 \{6 A_4 + A_1^2 A_2 [6 + 5 d (7d - 6)]\} \theta_0^8 \nonumber \\
&\hspace{85pt} + 512 A_1 A_2 d (19d-12) \theta_0^{10} \nonumber \\
&\hspace{85pt} + 4096 A_2 d^2 \theta_0^{12} \bigg\} \, . \nonumber
\end{align}
The magnification $\mu$ is expanded as
\begin{align}
\mu = \mu_0 + \mu_1 \epsilon + \mu_2 \epsilon^2 + \mu_3 \epsilon^3 + \mathcal{O}(\epsilon^4) \, ,
\end{align}
and the coefficients are
\begin{align}
\mu_0 &= \frac{16\theta_0^2}{A_1^2 - 16\theta_0^2} \, , \\
\mu_1 &= - \, \frac{16 A_2 \theta_0^3}{(A_1 + 4\theta_0^2)^3} \, , \\
\mu_2 &= \frac{8\theta_0^2}{3(A_1-4\theta_0^2)(A_1+4\theta_0^2)^5} \bigg\{ - A_1^6 d^2 + 8 A_1^2 [6 A_3 + A_1^3 (2 + 6 d - 9 d^2)] \theta_0^2 \nonumber \\
&\hspace{126pt} - 32 \{18 A_2^2 - 12 A_1 A_3 + A_1^4 [-4 + d (-12 + 17 d)]\} \theta_0^4 \\
&\hspace{126pt} - 128 [-6 A_3 + A_1^3 (-2 - 6 d + 9 d^2)] \theta_0^6 - 256 A_1^2 d^2 \theta_0^8 \bigg\} \, , \nonumber \\
\mu_3 &= \frac{8\theta_0}{3(A_1-4\theta_0^2)(A_1+4\theta_0^2)^7} \bigg\{ A_1^3 [6 A_2^3 - 12 A_1 A_2 A_3 + 6 A_1^2 A_4 + A_1^4 A_2 (2 - 3 d^2)] \\
&\hspace{126pt} + 4 A_1^2 (42 A_2^3 - 84 A_1 A_2 A_3 + 42 A_1^2 A_4 +  A_1^4 A_2 (14 + 3 (4 - 9 d) d)) \theta_0^2 \nonumber \\
&\hspace{126pt} - 96 A_1 (-21 A_2^3 + 42 A_1 A_2 A_3 - 18 A_1^2 A_4 +  A_1^4 A_2 (5 d^2 - 4)) \theta_0^4 \nonumber \\
&\hspace{126pt} - 128 (-105 A_2^3 + 150 A_1 A_2 A_3 - 66 A_1^2 A_4 + 2 A_1^4 A_2 (9 d^2 + 3 d - 8)) \theta_0^6 \nonumber \\
&\hspace{126pt} - 256 (120 A_2 A_3 - 78 A_1 A_4 + A_1^3 A_2 (3 d (23 d - 4) - 38)) \theta_0^8 \nonumber \\
&\hspace{126pt} - 3072 (-6 A_4 + A_1^2 A_2 (11 d^2 - 6)) \theta_0^{10} \nonumber \\
&\hspace{126pt} + 24576 A_1 A_2 (-2 + d) d \theta_0^{12} \bigg\} \, . \nonumber
\end{align}
Last, the total magnification is expanded as the sum
\begin{align}
\mu_\text{tot} = \mu_\text{tot,0} + \mu_\text{tot,1} \epsilon + \mu_\text{tot,2} \epsilon^2 + \mu_\text{tot,3} \epsilon^3 + \mathcal{O}(\epsilon^4) \, ,
\end{align}
with the coefficients
\begin{align}
\mu_\text{tot,0} &= \frac{\beta^2 + (A_1/2)}{\beta\sqrt{\beta^2 + A_1}} \, , \\
\mu_\text{tot,1} &= 0 \, , \\
\mu_\text{tot,2} &= \frac{1}{12\beta{(\beta^2+A_1)}^{5/2}} \bigg[ 9 A_2^2 + 2 A_1^3 (A_1 + \beta^2) (9 d^2 - 6 d - 2)  - 12 (A_1 + \beta^2) A_3 + 4 A_1^2 d^2 (A_1 + \beta^2) \beta^2 \bigg] \, , \\
%\mu_\text{tot,3} &= \frac{1}{12 A_1^4{(\beta^2+A_1)}^{7/2}} \bigg\{ \hspace{13pt} 3 A_1^3 \{35 A_2^3 - 60 A_1 A_2 A_3 + 24 A_1^2 A4 + 2 A_1^4 A_2 [2 + (2 - 3 d) d]\} \\
%&\hspace{100pt} + 2 A_1^2 [105 A_2^3 - 210 A_1 A_2 A_3 + 96 A_1^2 A_4 + A_1^4 A_2 (26 + 30 DD - 51 d^2)] \beta^2 \\
%&\hspace{100pt} + 4 A_1 \{42 A_2^3 - 84 A_1 A_2 A_3 + 42 A_1^2 A4 + A_1^4 A_2 [14 + 3 (4 - 9 d) d]\} \beta^4 \\
%&\hspace{100pt} + 8 [6 A_2^3 - 12 A_1 A_2 A_3 + 6 A_1^2 A_4 + A_1^4 A_2 (2 - 3 d^2)] \beta^6 \bigg \}
\mu_\text{tot,3} &= \frac{2}{3 A_1^4 \beta} \bigg[ -6 A_2^3 + 12 A_1 A_2 A_3 - 6 A_1^2 A_4 + A_1^4 A_2 (-2 + 3 d^2) \bigg] \, ,
\end{align}
which concludes our considerations.

\begin{table*}[!hb]
\begin{tabular}{lclclclclclclclclclc} \hline \hline
Abbr. &~& Name &~& $a_1$ &~& $a_2$ &~& $a_3$ &~& $a_4$ &~& $b_1$ &~& $b_2$ &~& $b_3$ &~& $b_4$  \\[3pt] \hline
&&&&&& \\[-8pt]
S && Schwarzschild && $1$ && $0$ && $0$ && $0$ && $1$ && $1$ && $1$ && $1$ \\[12pt]
H && Hayward && $1$ && $0$ && $0$ && $2\xi^2$ && $1$ && $1$ && $1$ && $1 - \tfrac14 \xi^2$  \\[12pt]
M && Minkowski core && $1$ && $\xi$ && $\tfrac12 \xi^2$ && $\tfrac16 \xi^3$ && $1$ && $1 - \tfrac12 \xi$ && $\displaystyle 1 - \xi + \tfrac18 \xi^2$ && $1 - \tfrac32 \xi + \tfrac12 \xi^2 - \tfrac{1}{48}\xi^3$ \\[12pt]
W && Simpson--Visser && $1$ && $0$ && $0$ && $0$ && $1$ && $1 + \tfrac14 \xi^2$ && $1 + \tfrac14 \xi^2$ && $\displaystyle 1 + \tfrac14 \xi^2 + \tfrac{1}{16}\xi^4$ \\[12pt]
\hline \hline
\end{tabular}
\caption{Parametrized post-Newtonian (PPN) expansion coefficients for the Schwarzschild metric, Hayward metric, Minkowski core metric, and Simpson--Visser metric, up to fourth order as defined in Eq.~\eqref{eq:ppn-def}, using the notation $\xi = \ell/(GM)$. Note that for the Simpson--Visser metric a transformation to a new radius variable is required, cf.~Eq.~\eqref{eq:c-trafo}, where we find $A(\tilde{r}) = 1-2GM/\tilde{r}$ and $\tilde{B} = \tilde{r}^3/[(\tilde{r}-2GM)(\tilde{r}^2-\ell^2)]$. Moreover, all of the above metrics feature $a_1 = b_1 = 1$.}
\label{table:ppn-coefficients}
\end{table*}

\section{Ray traced images}
\label{app:images}

In order to keep the main text of the article as uncluttered as possible, we only displayed a small selection of ray traced images. However, the reader may find it instructive to see more details on the renders, which is why we opted to include a small selection of images in this appendix. In Fig.~\ref{fig:lightcurve-overview} we show several lightcurve image sets for the validation, near-zone, and far-zone scenario, and we list the numerical parameters utilized for the lightcurves in Table~\ref{table:numerical-parameters}; for completeness we also depict the strong lensing of the accretion disk model in Fig.~\ref{fig:disk-overview}.

\begin{figure*}[!h]
\centering
\includegraphics[width=\textwidth]{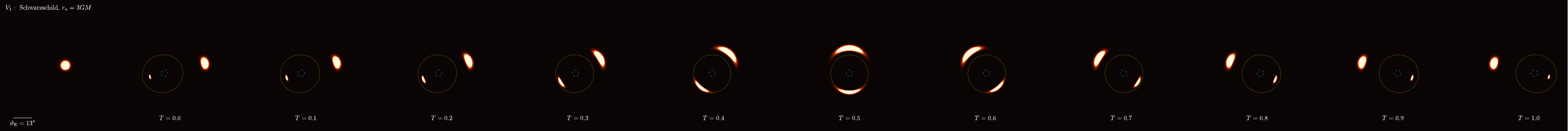} \\[10pt]
\includegraphics[width=\textwidth]{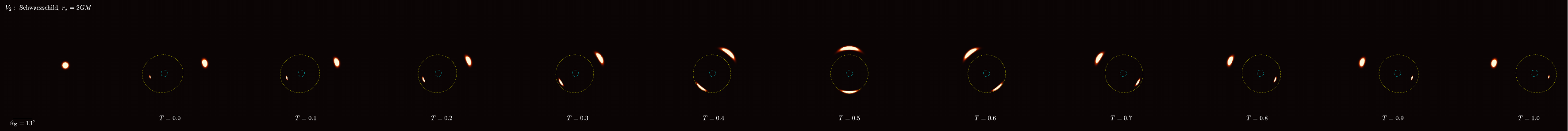} \\[10pt]
\includegraphics[width=\textwidth]{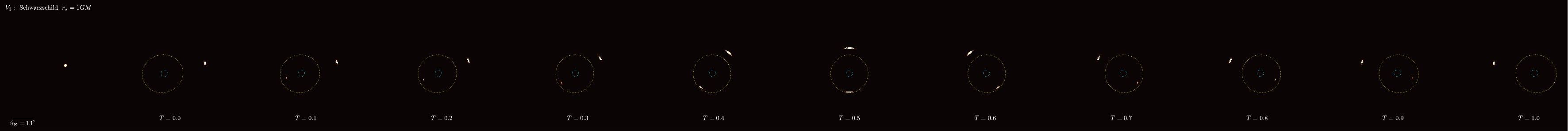} \\[10pt]
\includegraphics[width=\textwidth]{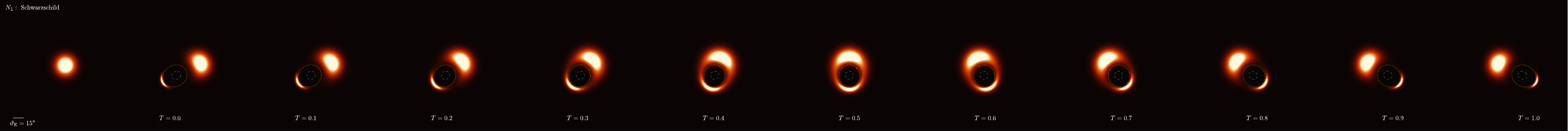} \\[10pt]
\includegraphics[width=\textwidth]{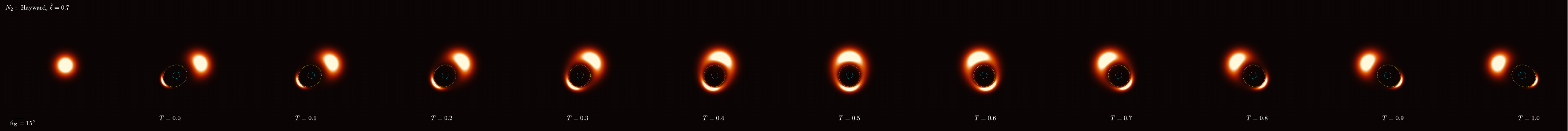} \\[10pt]
\includegraphics[width=\textwidth]{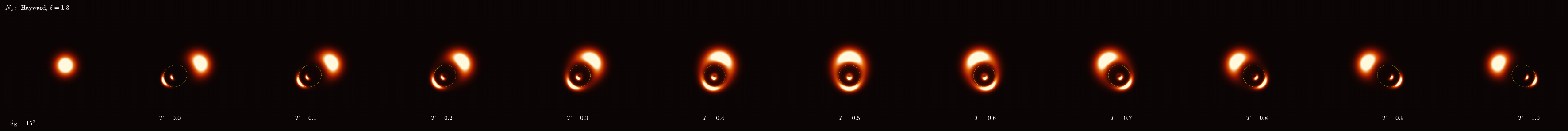} \\[10pt]
\includegraphics[width=\textwidth]{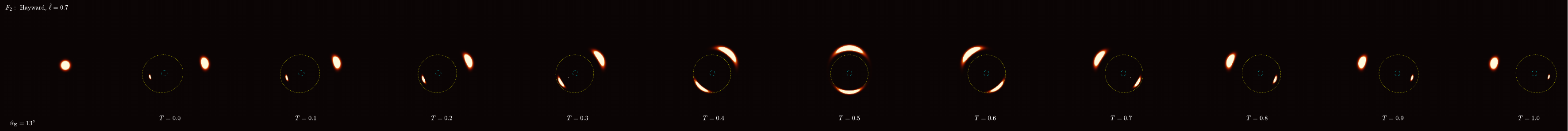} \\[10pt]
\includegraphics[width=\textwidth]{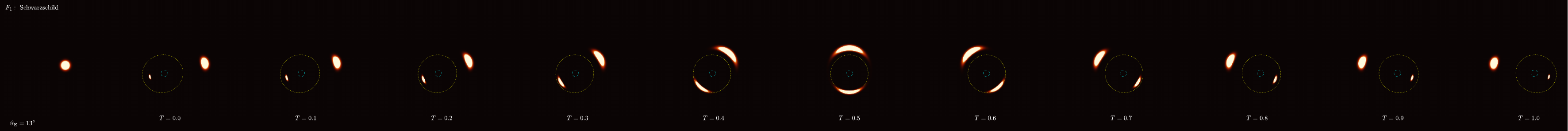} \\[10pt]
\includegraphics[width=\textwidth]{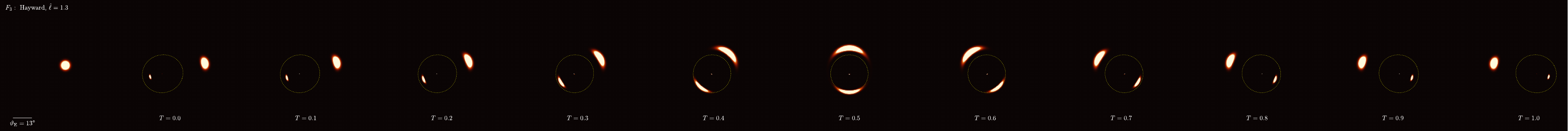}
\caption{Images used to extract the lightcurves in the validation scenario (top three rows, $V_1 \dots V_3$), the near-zone scenario (middle three rows, $N_1 \dots N_3$) and the far-zone scenario (bottom three rows, $F_1 \dots F_3$), normalized to the brightness of the star in the absence of the lens (first image in each row). Time steps go from $0.0$ (leftmost picture) to $1.0$ (rightmost picture) in steps of $0.1$. The simulations were performed with a time difference of $0.05$, but for the sake of illustration we only display every other image. The results shown (for the near-zone and far-zone scenario) cover only the $S$, $H_-$, and $H_+$ cases; the lightcurves for the $M_\pm$ and $W_\pm$ cases are qualitatively similar and hence omitted for brevity.}
\label{fig:lightcurve-overview}
\end{figure*}

\clearpage

\begin{table*}[!h]
\begin{tabular}{ccccccccccccccccccc} \hline \hline  \\[-10pt]
number &~& model &~& $\hat{\ell}$ &~& $r_\star~[GM]$ &~& $\omega~[GM]$ &~& $\alpha_\text{FOV}$ &~& $d_\text{ol}~[GM]$ &~& $d_\text{ls}~[GM]$ &~& $x_\perp~[GM]$ &~& $z_\perp~[GM]$ \\[3pt] \hline
&&  && && && \\[-8pt]
 $V_1$ && S       && $0.0$ && $3$ && $5$ && $6\vartheta_\text{E}$ && $50$ && $100$ && $20$ && $5$ \\
 $V_2$ && S       && $0.0$ && $2$ && $5$ && $6\vartheta_\text{E}$ && $50$ && $100$ && $20$ && $5$ \\
 $V_3$ && S       && $0.0$ && $1$ && $5$ && $6\vartheta_\text{E}$ && $50$ && $100$ && $20$ && $5$ \\[3pt] \hline \\[-8pt]
 $N_1$ && S       && $0.0$ && $3$ && $5$ && $8\vartheta_\text{E}$ && $20$ && $10$ && $10$ && $5$ \\ % 08
 $N_2$ && $H{}_-$ && $0.7$ && $3$ && $5$ && $8\vartheta_\text{E}$ && $20$ && $10$ && $10$ && $5$ \\ % 09
 $N_3$ && $H{}_+$ && $1.3$ && $3$ && $5$ && $8\vartheta_\text{E}$ && $20$ && $10$ && $10$ && $5$ \\ % 10
 $N_4$ && $M{}_-$ && $0.7$ && $3$ && $5$ && $8\vartheta_\text{E}$ && $20$ && $10$ && $20$ && $5$ \\ % 11
 $N_5$ && $M{}_+$ && $1.3$ && $3$ && $5$ && $8\vartheta_\text{E}$ && $20$ && $10$ && $20$ && $5$ \\ % 12
 $N_6$ && $W{}_-$ && $0.7$ && $3$ && $5$ && $8\vartheta_\text{E}$ && $20$ && $10$ && $20$ && $5$ \\ % 13
 $N_7$ && $W{}_+$ && $2.2$ && $3$ && $5$ && $8\vartheta_\text{E}$ && $20$ && $10$ && $20$ && $5$ \\[3pt] \hline \\[-8pt]  % 14
 $F_1$ && S       && $0.0$ && $3$ && $5$ && $6\vartheta_\text{E}$ && $50$ && $100$ && $20$ && $5$ \\ % 01
 $F_2$ && $H{}_-$ && $0.7$ && $3$ && $5$ && $6\vartheta_\text{E}$ && $50$ && $100$ && $20$ && $5$ \\ % 02
 $F_3$ && $H{}_+$ && $1.3$ && $3$ && $5$ && $6\vartheta_\text{E}$ && $50$ && $100$ && $20$ && $5$ \\ % 03
 $F_4$ && $M{}_-$ && $0.7$ && $3$ && $5$ && $6\vartheta_\text{E}$ && $50$ && $100$ && $20$ && $5$ \\ % 04
 $F_5$ && $M{}_+$ && $1.3$ && $3$ && $5$ && $6\vartheta_\text{E}$ && $50$ && $100$ && $20$ && $5$ \\ % 05
 $F_6$ && $W{}_-$ && $0.7$ && $3$ && $5$ && $6\vartheta_\text{E}$ && $50$ && $100$ && $20$ && $5$ \\ % 06
 $F_7$ && $W{}_+$ && $2.2$ && $3$ && $5$ && $6\vartheta_\text{E}$ && $50$ && $100$ && $20$ && $5$ \\[3pt]   \hline \hline % 07
\end{tabular}
\caption{We list the numerical parameters employed in the various lightcurves discussed in this paper. $V_1 \dots V_3$ are used for the validation of the numerical finite-size setup against the theoretical point-source formulae; $N_1 \dots N_7$ comprise the near-zone configurations, while $F_1 \dots F_7$ are the far-zone configurations.}
\label{table:numerical-parameters}
\end{table*}

\clearpage

\begin{figure*}[!ht]
\centering
\includegraphics[width=0.18\textwidth]{data/disk01/disk01-dat-inc=0-c=2.5-arc=29.59.pdf} \hspace{5pt}
\includegraphics[width=0.18\textwidth]{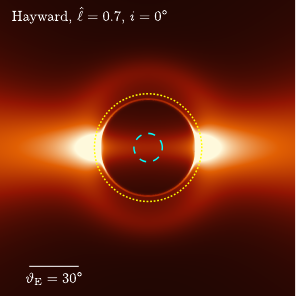} \hspace{5pt}
\includegraphics[width=0.18\textwidth]{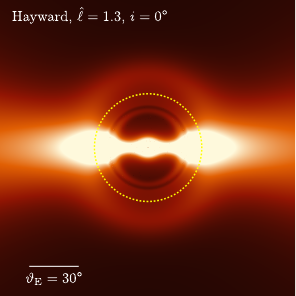} \hspace{5pt}
\includegraphics[width=0.18\textwidth]{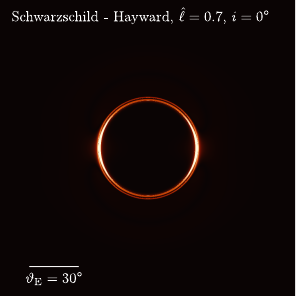} \hspace{5pt}
\includegraphics[width=0.18\textwidth]{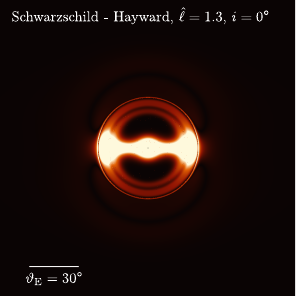} \\[8pt]
\includegraphics[width=0.18\textwidth]{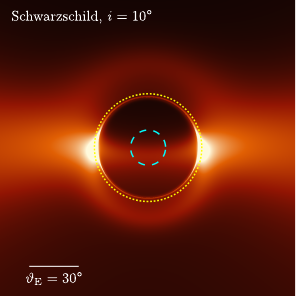} \hspace{5pt}
\includegraphics[width=0.18\textwidth]{data/disk01/disk02-dat-inc=10-c=2.5-arc=29.59.pdf} \hspace{5pt}
\includegraphics[width=0.18\textwidth]{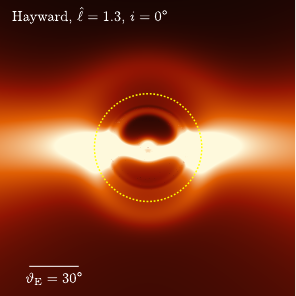} \hspace{5pt}
\includegraphics[width=0.18\textwidth]{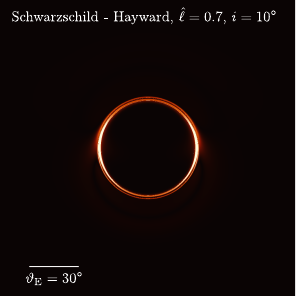} \hspace{5pt}
\includegraphics[width=0.18\textwidth]{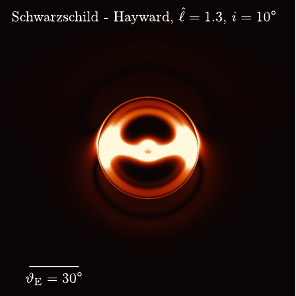} \\[8pt]
\includegraphics[width=0.18\textwidth]{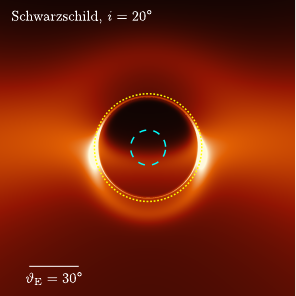} \hspace{5pt}
\includegraphics[width=0.18\textwidth]{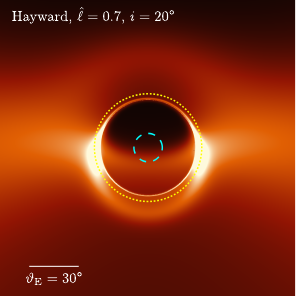} \hspace{5pt}
\includegraphics[width=0.18\textwidth]{data/disk01/disk03-dat-inc=20-c=2.5-arc=29.59.pdf} \hspace{5pt}
\includegraphics[width=0.18\textwidth]{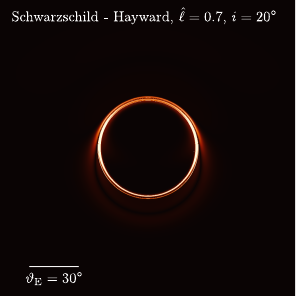} \hspace{5pt}
\includegraphics[width=0.18\textwidth]{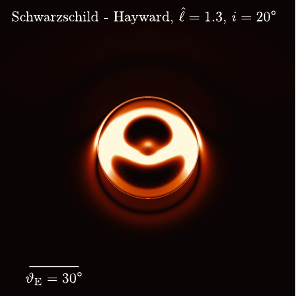} \\[8pt]
\includegraphics[width=0.18\textwidth]{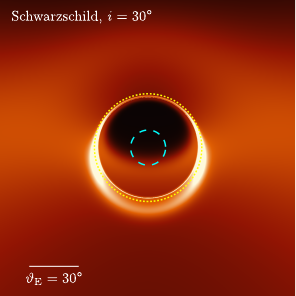} \hspace{5pt}
\includegraphics[width=0.18\textwidth]{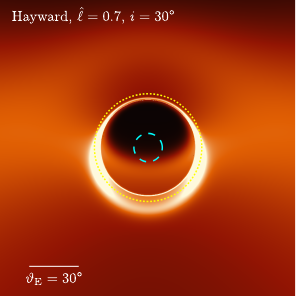} \hspace{5pt}
\includegraphics[width=0.18\textwidth]{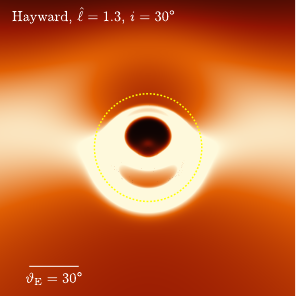} \hspace{5pt}
\includegraphics[width=0.18\textwidth]{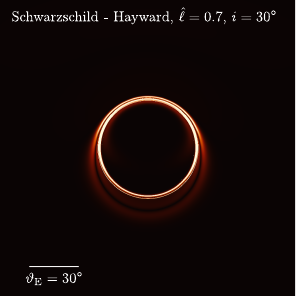} \hspace{5pt}
\includegraphics[width=0.18\textwidth]{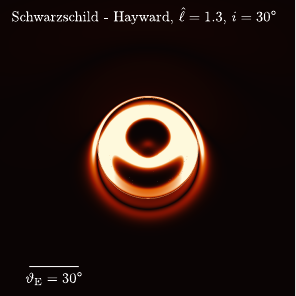} \\[8pt]
\includegraphics[width=0.18\textwidth]{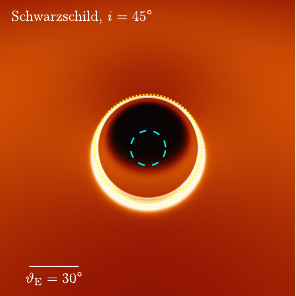} \hspace{5pt}
\includegraphics[width=0.18\textwidth]{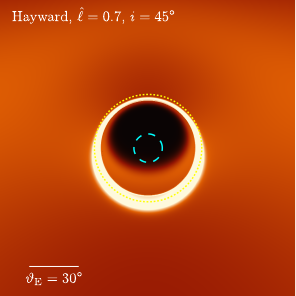} \hspace{5pt}
\includegraphics[width=0.18\textwidth]{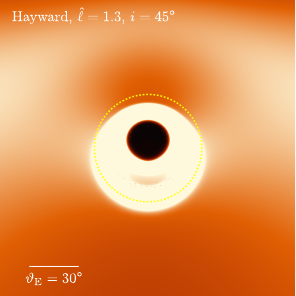} \hspace{5pt}
\includegraphics[width=0.18\textwidth]{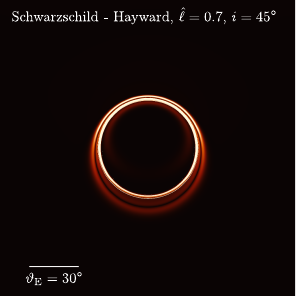} \hspace{5pt}
\includegraphics[width=0.18\textwidth]{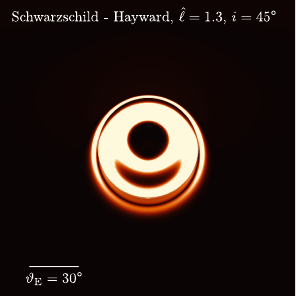} \\[8pt]
\includegraphics[width=0.18\textwidth]{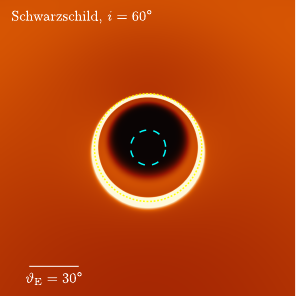} \hspace{5pt}
\includegraphics[width=0.18\textwidth]{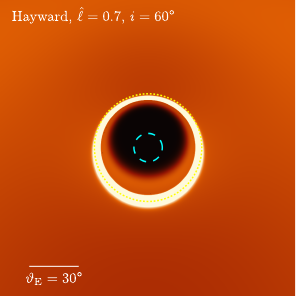} \hspace{5pt}
\includegraphics[width=0.18\textwidth]{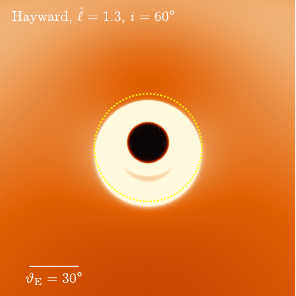} \hspace{5pt}
\includegraphics[width=0.18\textwidth]{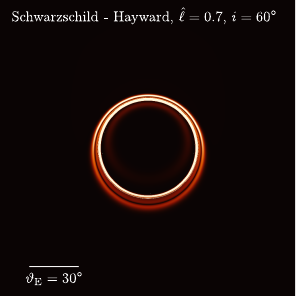} \hspace{5pt}
\includegraphics[width=0.18\textwidth]{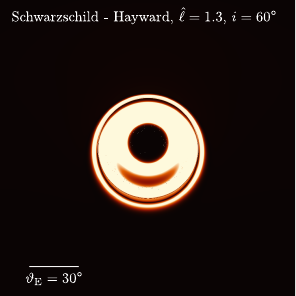} 
\caption{Accretion disk luminosity for the scenarios $S$, $H_-$ and $H_+$ at different inclinations ($i=0^\circ$ corresponds to an edge-on view, and $i>0$ describes the angle out of the equatorial plane). The fourth and fifth columns show the absolute value of the differences $|S-H_-|$ and $|S-H_+|$, respectively. The Einstein angle is visualized at the bottom left, the dashed cyan line is the projected outline of the black hole horizon (if present), and the yellow dashed line is the projected Einstein radius. The images are individually normalized to their brightest spot and then amplified by a factor of 2.5 to enhance contrast. While the morphology of the accretion disk luminosity is qualitatively similar in the $S$ and $H_-$ cases, the absence of a horizon in the $H_+$ scenario leads to noticeable differences.}
\label{fig:disk-overview}
\end{figure*}

\clearpage

\end{widetext}

\end{document}